\documentclass[%
preprint,
 amsmath,amssymb,
 aps,
floatfix,
showkeys,
]{revtex4-2}

\usepackage{graphicx}
\usepackage{dcolumn}
\usepackage{bm}
\usepackage{xcolor}
\usepackage{placeins}

\begin{document}


\preprint{APS/123-QED}

\title{Correlation of viral loads in disease transmission chains could bias early estimates of the reproduction number}

\author{Thomas Harris}
\author{Nicholas Geard}
\author{Cameron Zachreson}
\affiliation{%
 School of Computing and Information Systems, The University of Melbourne, Parkville, Victoria, Australia.
}%

\date{\today}

\begin{abstract}
Early estimates of the transmission properties of a newly emerged pathogen are critical to an effective public health response, and are often based on limited outbreak data. Here, we use simulations to investigate a potential source of bias in such estimates, arising from correlations between the viral load of cases in transmission chains. We show that this mechanism can affect estimates of fundamental transmission properties characterising the spread of a virus. Our computational model simulates a disease transmission mechanism in which the viral load of the infector at the time of transmission influences the infectiousness of the infectee. These correlations in transmission pairs produce a population-level decoherence process during which the distributions of initial viral loads in each subsequent generation converge to a steady state. We find that outbreaks arising from index cases with low initial viral loads give rise to early estimates of transmission properties that are subject to large biases. These findings demonstrate the potential for bias arising from transmission mechanics to affect estimates of the transmission properties of newly emerged viruses. 
\end{abstract}
\maketitle

\section{Introduction}

Infectious disease outbreaks have dramatic impacts on communities. Accurate estimates of a pathogen's transmission properties, such as the serial interval and basic reproduction number, are critical to inform the design of appropriate control measures. These properties are typically estimated based on outbreak data; therefore, estimation for a newly emerged pathogen can be challenging because limited data is available \cite{mercer2011effective}. Estimates of these quantities are subject to multiple potential biases, with some biases stemming from the methods used to collect outbreak data and others arising from the statistical methods used for parameter estimation  \cite{mercer2011effective,thompson2019improved}. Failure to account for these biases can lead to inaccurate conclusions around the scale of the threat posed by a particular pathogen, and result in subsequent public health responses that are poorly calibrated to the risk a community faces.
\\\\
The basic reproduction number, $R_0$, is defined as the average number of secondary cases produced by a single {\it{typical}} infected individual in an otherwise susceptible population. The effective reproduction number, $R_t$, is the time-dependent number of secondary cases produced by a typical infectious case at time $t$ after the introduction of the pathogen. Theoretically, the parameters $R_0$ and $R_t$ are approximately equivalent during the early exponential growth of a pathogen within a large population of susceptible hosts. Numerous methods have been developed for estimating $R_t$ at some time $t$ in an ongoing outbreak \cite{thompson2019improved,gostic2020practical,cori2013new,wallinga2004different,green2022inferring}, and these are often used to estimate $R_0$ based on early outbreak data. Cori {\it et al.} \cite{cori2013new} describe a method for estimating $R_t$, that uses infection incidence (number of new detected infections per day) up to some time $t$ and knowledge of the serial interval (the time between the onset of symptoms in an infector and infectee transmission pair). This approach has been used in the analysis of various outbreaks \cite{ali2013transmission,ferguson2016countering,zhang2021transmission}, and has been extended to account for heterogeneity in transmission, capturing differences in transmission potential between a discrete set of groups in a population \cite{green2022inferring}.
\\\\
Several studies have investigated potential sources of bias that can affect estimates of the reproduction number \cite{mercer2011effective, thompson2019improved, gostic2020practical}. For example, incorrectly including imported cases in local case counts exaggerates the number of secondary cases attributed to local spread, and can lead to overestimates of the transmissibility of a pathogen \cite{mercer2011effective,thompson2019improved}. Similarly, these studies identified over-representation of lower or higher transmission events early in an outbreak \cite{mercer2011effective} and appropriate selection of serial interval data \cite{thompson2019improved} as important potential sources of bias to consider in establishing accurate estimates of the reproduction number. While these ``experimental" sources of bias can be accounted for through enhanced surveillance, another complicating factor is the natural heterogeneity in the disease progression of each individual. For example, diverse expressions of illness in infected individuals can affect the capacity to understand transmission of a disease \cite{green2022inferring}. 
\\\\
For viruses, the viral load dynamics within hosts have been found to vary between infected individuals. The reason for this heterogeneity in viral load dynamics is not well understood. One possible mechanism that could contribute is variation in the initial viral load, the viral quantity transmitted from a donor host that initiates an infection in an exposed recipient \cite{li2014modeling,chu2004initial,guallar2020inoculum, prince1993pathogenesis, ottolini1996semi, liu2009primary}. Key aspects of viral load dynamics, such as duration of infection and peak viral load, are impacted by the initial viral load \cite{callison2006development,powell2006immune}. Furthermore, a positive relationship has been observed between a host’s viral load and their infectiousness \cite{handel2015crossing,marks}. From the modelling perspective there has been discussion on the strong intuitive connection between pathogen load and infectiousness, while also acknowledging the possible complications introduced by varying symptom development and host behaviour \cite{handel2015crossing}. 
\\\\
In this work, we form and investigate a hypothesis based on the well-established findings discussed above, namely that:
\begin{enumerate}
    \item higher viral load corresponds to increased probability of transmission to a new susceptible host (given contact), and
    \item being exposed to higher levels of virus produces higher peak viral loads because of the larger initiating quantity.
\end{enumerate}
Intuitively, these two results indicate the potential for correlations to exist in transmission chains, i.e. that a transmission chain initiated with a highly infectious index case may generate higher-than-average transmission rates, and vice versa. Our hypothesis is that such correlations of within-host viral dynamics between transmission pairs could bias estimates of transmission properties produced during the early stages of an outbreak.
\\\\
To investigate this hypothesis, we designed and implemented a multiscale model of infectious disease transmission to capture correlations of within-host viral load dynamics and their effects on transmission property estimation. Multiscale models have increasingly been used to capture infectious disease spread \cite{childs2019linked,mideo2008linking,hart2020theoretical}. These models typically capture disease dynamics at both a within-host scale (i.e. how an infection progresses inside a single individual)  and a between-host scale (i.e. how the infection is transmitted between multiple individuals of a host population) \cite{childs2019linked}. These types of models have been used to help address open questions relating to a variety of infectious diseases, such as Influenza A \cite{handel2013multi} and Ebola \cite{Nguyen2018}. Importantly, multiscale modelling allows for explicit representation of individual-level processes, such as host viral load, and the linking mechanism that connects these processes to population-level spread dynamics. 
\\\\
Our model describes a general acute respiratory virus spreading in a population. We assume three key relationships in the transmission of the virus:
\begin{enumerate}
    \item Host infectiousness increases with viral load
    \item The trajectory of the host viral load over the course of infection is related to the initiating quantity of virus
    \item Recipient initial viral load increases with the donor viral load at the time of transmission \\
\end{enumerate}
Assumption 1 is supported by experimental studies such as \cite{marks} and has been used in previous simulation studies \cite{handel2015crossing,hart2020theoretical, chen2009viral}. Similarly, assumption 2 is supported by experimental studies \cite{callison2006development,powell2006immune} and several existing within-host models of pathogens \cite{li2014modeling}. While we are not aware of any direct evidence supporting assumption 3, this assumption is consistent with studies of SARS \cite{subbarao2004prior} and tuberculosis \cite{saini2012ultra} that suggest the inoculum dose size (the quantity of pathogen presented to a susceptible host), has a positive relationship with the pathogen load measured after infection. Assumption 3 has also been used by previous simulation studies \cite{steinmeyer2010methods,Nguyen2018}.
\\\\
We investigate how these widely-held assumptions, related to the mechanisms underpinning viral transmission, introduce a source of bias into early outbreak estimates. We found correlations in transmission pairs produce a population-level decoherence process, during which the distributions of initial viral loads in each subsequent generation converge to a steady state. We also found that outbreaks arising from index cases with low initial viral loads give rise to early estimates of transmission properties that are subject to large biases. Our work demonstrates how biologically feasible assumptions around viral transmission can introduce complex population-level biases in estimations of disease spread.

\section{Methods}

In this section we describe: 
\begin{itemize}
    \item{the multiscale model used to simulate disease spread in a host population,}
    \item{the analytical method we developed to quantify the rate at which correlations in viral load dissipate over the course of an outbreak and,}
    \item{the methods we used to estimate reproduction numbers and serial intervals based on simulated outbreak data.}
\end{itemize}
At the end of the section, we outline the three main experiments conducted in this study.

\subsection{Multiscale model of disease dynamics}

\begin{figure}[p!]
    \centering
    \includegraphics[scale=0.6]{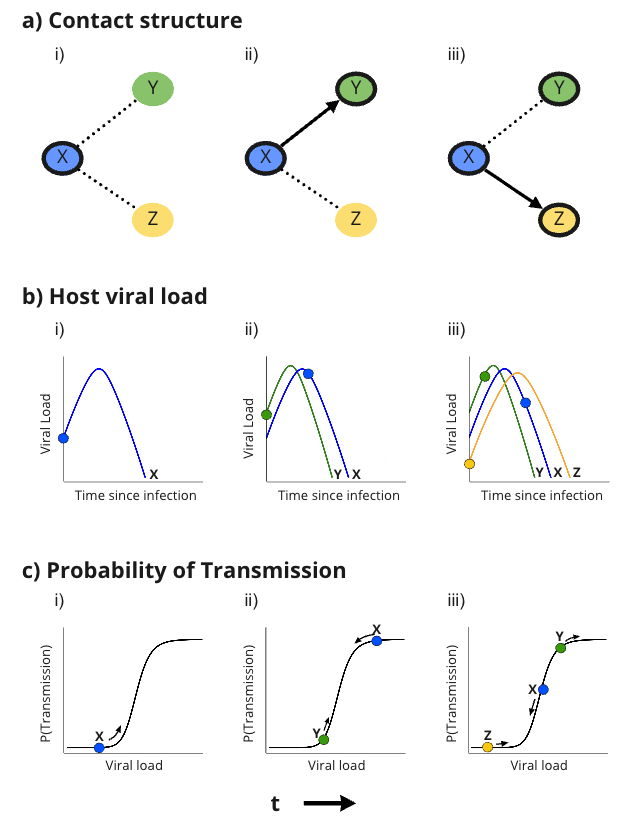}
    \caption{Schematic of the mechanistic model of disease transmission. The viral load trajectory of each secondary case depends on the viral load of the primary case at the time of transmission. Here, the primary case X (i) infects two secondary cases, Y (ii) and Z (iii). The viral load trajectories of cases Y and Z depend on the viral load of the primary case X at the time of transmission (b), which also influences the probability of transmission occurring via the sigmoidal mapping in (c).}
    \label{fig:fig1}
\end{figure}

\subsubsection{Model of case progression and virus transmission between hosts}
Between-host disease transmission is described by an agent-based model (ABM). This modelling approach is well suited to capturing infectious disease spread, because it allows for the explicit simulation of complex interaction behaviour and heterogeneity between individuals. ABMs are used across the infectious disease multiscale modelling literature \cite{childs2019linked}.
\\\\
Our agent-based model simulates random interactions between individuals in a large, well-mixed population. At any point in time, an agent exists in one of four states, which correspond to the compartments of a Susceptible-Exposed-Infectious-Recovered (SEIR) compartmental model. At each timestep, infectious individuals are randomly paired with a single susceptible host. Transmission may then occur, with the probability of transmission determined as a function of the viral load of the infector (Figure \ref{fig:fig1}). If transmission occurs, the within-host model of viral load as a function of time determines the incubation period (time spent in the ``exposed" state), the infectious period, and the time until recovery. There are twelve time steps ($S_d=12$) in each simulation day (i.e. interactions between individuals are simulated 12 times across a day).
\\\\
To compute the viral load as a function of time since infection, we define a within-host model of disease progression (see Section \ref{WH_model}) which guides transitions between compartments. As mentioned above, infected individuals are initially ``exposed", that is, they cannot transmit infection. As their viral load increases past a defined threshold $V_T$ (here set to $V_T = 13431.67$---see Supplementary Material \ref{sup1}), they transition from the ``exposed" to ``infectious" state, where they remain for the duration of their infection. After passing through a peak, the infected individual's viral load gradually decreases. When the viral load decreases past $V_T$, the individual is no longer infectious, and has transitioned to the ``recovered" state. 

\subsubsection{Model of within-host viral dynamics}

\label{WH_model}

To simulate viral load dynamics within individual hosts, we chose a mathematical model developed by Steinmeyer {\it et al.} \cite{steinmeyer2010methods} for the dynamics of acute respiratory infections similar to those caused by Influenza or SARS. These types of models have been applied extensively to within-host viral dynamics, and typically include features associated with self-limiting viral replication as well as the activity of the host immune response \cite{li2014modeling, baccam2006kinetics, handel2010towards,  steinmeyer2010methods}.
\\\\ 
The system of ordinary differential equations (ODEs) that describe the within-host viral dynamics is given by:
\begin{equation}
    \frac{dV}{dt} = rV - V(\frac{rV}{K_v}+k_{I}I+k_{N}N+k_{P}P) \,,
\end{equation}
\begin{equation}
    \frac{dI}{dt} = a_{I}V+b_I(1-\frac{I}{K_I}) \,,
\end{equation}
\begin{equation}
    \frac{dN}{dt} = a_{N}V\Theta(t-\tau_{N})-d_{N}N \,,
\end{equation}
\begin{equation}
    \frac{dP}{dt} = a_{P}VP+cN(t-\tau_P) \,,
\end{equation}
where $V$ is the number of viral particles, $I$ is the strength of the innate immunity, $N$ is the number of non-specific memory cells, $P$ is the number of specific memory cells, $t$ is the time since initial infection, $r$ is the viral replication rate, $K$ is the carrying capacity (population of each cell type, denoted by subscript letter, that can be supported in the body), $a$ is the growth rate per virion relevant to each immune defense (denoted by subscript letter), $k$ is the rate of viral removal relevant to each immune defense (denoted by subscript letter),  $\tau$ represents the time delay after which the development of non-specific memory cells ($\tau_{N}$) or specific memory cells ($\tau_{P}$) can occur, $b_I$ is the constant rate of growth for the innate immunity, $c$ is the specific memory cell growth rate proportional to non-specific memory cells, and $d_N$ is the non-specific memory cell decay rate. $\Theta$ describes a Heaviside step function given by:
\begin{equation}
    \Theta(t) = \begin{cases} 
                      0, & \text{if t $<$ 0} \\
                      1, & \text{if t $\geq$ 0}
                 \end{cases} 
\end{equation}
The specific parameter values chosen in our model as-implemented are provided in the Supplementary Material (see Supplementary Material \ref{sup1}). To introduce individual-level heterogeneity of within-host immune response, carrying capacity parameters ($K_v$ \& $K_I$) and the rates of viral removal relevant to each immune defense ($k_I$, $k_N$ \& $k_P$) were sampled from gamma distributions for each infected individual (see Supplementary Material \ref{sup1}). Taken together, this choice of model and accompanying parameters describe within-host viral load dynamics that depend substantially on the initial viral load (see Supplementary Material \ref{sup2}). Specifically, higher initial viral loads correspond with a faster viral proliferation and a higher peak load, but also produce a stronger immune response and a faster recovery. See \cite{steinmeyer2010methods} for a more detailed description of the within-host model used here.

\subsubsection{Modelling correlations in viral load between cases in transmission chains}
For each interaction between an infectious donor and susceptible recipient host, the donor viral load at transmission determines two key values: infectiousness of the donor and the initial viral load of the recipient. At the time of exposure, the donor viral load ($V_d$) translates to a probability of transmission, given exposure ($P_{trans}$) {\it via} a sigmoidal mapping (Figure \ref{fig:fig1}). This mapping is given by:
\begin{equation}
\label{ptrans}
    P_{trans} = \frac{c_1V_{d}^{\zeta}}{V_{d}^{\zeta}+c_2^{\zeta}} \,,
\end{equation}
where $c_{1}$ is the maximum $P_{trans}$, $c_{2}$ is the donor viral load at which $P_{trans}$ is half-maximal, and $\zeta$ is a slope parameter characterising the steepness of the transition from low to high infectiousness. The specific parameter values chosen in our model as-implemented are provided in the Supplementary Material (see Supplementary Material \ref{sup3}). 
The sigmoidal shape was chosen based on experimental results suggesting this kind of relationship may describe transmission likelihood \cite{handel2015crossing,saini2012ultra}, and its use in other models of disease transmission \cite{hart2020theoretical,henriques2022modelling}.
\\\\
Second, given transmission has occurred, the donor viral load ($V_d$) translates to a value for the initial viral load that initiates infection in the recipient ($V_R(0)$) also {\it via} a sigmoidal mapping:
\begin{equation}\label{eq_V0}
    V_R(0) = \frac{d_1V_d^{\kappa}}{V_d^{\kappa}+d_2^{\kappa}}\,,
\end{equation}
where $d_{1}$ is the maximum $V_R(0)$, $d_{2}$ is the donor viral load at which $V_R(0)$ is half-maximal, and $\kappa$ is a slope parameter characterising the steepness of the transition from low to high $V_R(0)$. The specific parameter values chosen in our model as-implemented are provided in the Supplementary Material (see Supplementary Material \ref{sup4}). Though it is not supported by any direct empirical evidence we are aware of, we chose the sigmoid function in Equation \ref{eq_V0} because there is evidence suggesting this functional form could likely describe the relationship between pathogen load and the likelihood of transmission , a closely related property \cite{handel2015crossing,saini2012ultra}.
\subsection{Quantifying the convergence of transmission dynamics}
\label{decoher}

In this model of disease transmission, the initial viral load of a host determines the trajectory of their infectiousness over time. Therefore, at any time $t$ during a simulated outbreak, the distribution of initial viral loads $P(V(0)~|~t)$ of the cases currently infected provides a useful summary statistic for the underlying dynamics. By examining how $P(V(0))$ changes as a function of time, we observe {\it decoherence}, that is, the initial state evolves until the transmission dynamics converge to a stable condition that does not depend on the initial state. The time required for convergence is a decreasing function of the viral load of the index case. This occurs because our transmission model produces correlation between the viral load of an infectious individual at the time of transmission and the initial viral load of the secondary case. 
\\\\
To quantify this decoherence process, we define a tolerance of $\delta = 0.0475$ and perform a Kolmogorov-Smirnov (KS) test to compute the KS statistic comparing  $P(V(0)~|~g)$ for each subsequent generation $g$ of infectious cases (note that here we compute the distribution of initial viral loads for each generation, rather than for a snapshot in time). After the generation $g_c$ when the KS statistic $KS(P(V(0)|g_c),P(V(0)|g_c + 1)) < \delta$ and $KS(P(V(0)|g_c),P(V(0)|g_c + 2)) < \delta$, we consider the system's dynamics to be stable. Stability means that the properties of the transmission dynamics will no longer change substantially as time goes on, as long as the infected and recovered populations are much smaller than the susceptible population.

\subsection{Estimation of transmission properties}
\label{est}
To understand specifically how the transmission process of this model could introduce a bias into early outbreak data, we investigated how the initial viral load of the index case ($V(0)_{index}$) affected estimation of the serial interval and the reproduction number
across the early portion of an outbreak. The methods outlined in this section are further described in the Supplementary Material (see Supplementary Material \ref{sup5}).
\subsubsection{Estimating the serial interval}
Because our model does not explicitly represent symptom expression, we used the time between infection and peak viral load as a surrogate for the incubation period (the time between infection and symptom onset). We assumed that symptom onset was always observed and reported on the day at which symptom onset occurred. We also assumed all transmission pairs could be established immediately without error, once the infectee symptom onset had occurred. Serial intervals were then determined by measuring the difference between symptom onset times of donor and recipient pairs.
\\\\
As each outbreak continued to grow after each observation time, we accounted for the effects of right truncation in the data \cite{nishiura2020serial}. We corrected for these effects by limiting the data to transmission pairs where 1) the donor is no longer infectious, and 2) all secondary cases produced by the donor have reached symptom onset.
\\\\
We fitted a gamma distribution to the set of serial intervals observed. We then performed parametric bootstrapping to estimate the uncertainty associated with the fitted distribution. This process generated a set of $N_B=100$ bootstrapped instances (sets of shape ($\alpha$) and rate ($\beta$) parameters). From this set, we were able to establish the mean and 95\% confidence intervals of the fitted gamma parameters. We used this bootstrapping approach to characterise the uncertainty in the serial interval and propagate this uncertainty to the reproduction number estimation process, as has been done effectively using Markov Chain Monte Carlo (MCMC) methods elsewhere \cite{thompson2019improved}.

\subsubsection{Estimating $R_0$}

The methods for estimating the reproduction number often target $R_t$---the average number of secondary cases produced by an infected individual in a population that may have some level of immunity, at some time $t$ in an ongoing outbreak 
\cite{thompson2019improved,gostic2020practical,cori2013new,wallinga2004different,green2022inferring}. Using our model, we simulated an early outbreak time period where depletion of susceptible individuals is yet to occur, and individuals have no existing immunity. With these assumptions, we estimated the basic reproduction number ($R_0$) by estimating $R_t$ during the early exponential phase of outbreak growth.
\\\\
After estimating the serial interval distribution for a particular outbreak, we used a method to estimate $R_0$ similar to the Bayesian technique developed by Cori {\it et al.} \cite{cori2013new}.
\\\\
The process for estimation of $R_0$ at some time $t$ in a given outbreak can be described by four key steps:
\begin{itemize}
    \item For each bootstrapped instance of the serial interval distribution describing this outbreak, we discretised the continuous gamma distribution. This discretisation produced a discrete probability mass function $\omega_s$, which described the likelihood of the serial interval being some integer number of days $s$.
    \item We computed the expectation of infectivity ($\Lambda_j$) for each day $j$ in the the period of days $t-\tau$ to $t$, where we assumed the reproduction number remains constant over this time period. The expectation of infectivity ($\Lambda_j$) on some day $j$ is computed as:
    \begin{equation}
        \Lambda_j = \sum_{s=1}^{j}I_{j-s}\omega_s \,,
    \end{equation}
    where $I_{j-s}$ refers to case incidence on day $j-s$ and  $\omega_s$ refers to the discrete probability mass function.
    \item We used the Bayesian framework detailed in \cite{cori2013new} to describe the posterior distribution of the reproduction number at time $t$ as a gamma distribution with shape ($\alpha$) and scale ($\beta$) parameters:
    \begin{equation}
        \alpha = a + \sum_{k=t-\tau}^{t}I_k \,,
    \end{equation}
    \begin{equation}
        \beta = \frac{1}{\frac{1}{b}+\sum_{k=t-\tau}^{t}\Lambda_k} \,,
    \end{equation}
    where $a$ and $b$ refer to the shape and scale parameters, respectively, of the assumed gamma prior distribution. Here, we chose $a$ and $b$ so that the mean and standard deviation of the prior distribution are both equal to five ($a=1$ \& $b=5$), similar to other studies \cite{cori2013new,thompson2019improved}. These parameter values ensured the prior distribution was relatively uninformative and conservative.
    \item Finally, we sampled from the defined posterior distribution $N_S=100$ times to produce a set of estimates of the reproduction number at time $t$. We pooled over all the sets of $R_0$ estimates produced from the bootstrapped instances of the serial interval distribution, establishing a set of $N_S \times N_B$ estimates of $R_0$ at time $t$ in a given outbreak.
\end{itemize}
See \cite{cori2013new} for a more detailed description of the Bayesian framework used here.
\\\\
Because we used a stochastic model of disease transmission, we simulated multiple outbreaks for each scenario to characterise the typical behaviour of the system for each set of conditions. To analyse a set $O$ of outbreaks for a scenario, we computed the mean $R_0$ estimate across $O$ for each day $d$ after the first case is detected. Specifically, we pooled the $N_S \times N_B$ $R_0$ estimates produced from each outbreak simulation, observed for $d$ days after the first case, to create a set of $R_0$ estimates of size $|O|\times N_S \times N_B$.
We then computed the mean of this set of $R_0$ estimates as a global summary statistic, 
which we denote as $\langle R_{0} \rangle_{d}$.


\subsubsection{Measuring $R_0$ estimation bias and confidence}
We measured the magnitude of the bias and the certainty associated with the 
distribution of $R_0$ estimates produced at each day $d$ in a single outbreak simulation, $i$, in a set of outbreak simulations, $O$. 
\\\\
To measure the bias, we computed how the maximum likelihood estimate of the distribution computed using observations up to day $d$ compares to some approximation of the `true' $R_0$ which we approximate as the mean over the set of outbreaks $O$ of $R_0$ estimates made on day 50:
\begin{equation}
    Bias(i,d)=\frac{|\langle R_{0}\rangle_{i,d} - R_{0}^{true}|}{R_{0}^{true}} \,,
\end{equation}
where $\langle R_{0}\rangle_{i,d}$ is the maximum-likelihood $R_0$ estimate on day $d$ for simulation $i$, and 
\begin{equation}
    R_{0}^{true} = \frac{\sum_{i=1}^{|O|}\langle R_{0}\rangle_{i,50}}{|O|} = \langle R_{0} \rangle_{50}
\end{equation}
We note here that approximating $R_{0}^{true}$ using the day 50 estimates was based on our knowledge of the distribution of generation intervals (the length of time between donor and recipient infections) that are produced in the model, and the number of generations typically required for the system's dynamics to converge ($g_c$). Specifically, the mean serial interval (which we use to approximate the generation interval) was known to be $\approx 4.2$ days and $g_c$ was seen to be at most 6--7 generations; therefore, observations at day 50 are well beyond the typical range of days for which system dynamics will converge (i.e. $4.2\times7 = 29.4 < 50$). 
\\\\
We also measured the relative confidence in each distribution by computing the ratio of the maximum likelihood estimate to the width of the 95\% confidence interval:
\begin{equation}
    Confidence(i,d)=\frac{\langle R_{0}\rangle_{i,d}}{q_{0.975}-q_{0.025}} \,,
\end{equation}
where $\langle R_{0}\rangle_{i,d}$ is the maximum-likelihood $R_0$ estimate on day $d$ for simulation $i$, and $q_{0.975}-q_{0.025}$ is width of the 95\% confidence interval. 
\\\\
To summarise the bias and confidence for some day $d$ across a simulated set of outbreaks, $O$, we computed:
\begin{equation} \label{mean_bias}
    Bias'(d)=\frac{\sum_{i=1}^{|O|}Bias(i,d)}{|O|} \,,
\end{equation}
\begin{equation} \label{mean_conf}
    Confidence'(d)=\frac{\sum_{i=1}^{|O|}Confidence(i,d)}{|O|} \,,
\end{equation}

\subsection{Experimental design}

\subsubsection{Effect of the initial viral load of the index case on outbreak growth}
To understand how the initial viral load of the index case affected early outbreak development, we varied $V(0)_{index}$ and measured the effect on case incidence during the early stages of outbreaks. We investigated three initial viral load values---$V(0)_{index}=4.5, 45$ \& $450$---and initialised 200 outbreaks with each setting. This number of simulations for each $V(0)_{index}$ setting ensured we accurately captured the range of behaviour that could be produced under each setting. For each simulated outbreak, we measured the case incidence over the first 50 days after the first detected case. We disregarded simulations in which the outbreak dies out (i.e. case prevalence reaches 0) before 50 days. Cases were detected perfectly at symptom onset (approximated by the time of peak viral load, $t_{Vmax}$). We scaled the sigmoid function describing transmission potential as a function of viral load, $P_{trans}$ (see Equation \ref{ptrans}), such that the $R_{0}^{true} \approx 2$ (see Supplementary Material \ref{sup3}). This scaling of the transmission rate meant that simulating uninhibited growth across all index case viral loads was computationally feasible. Note that this differs from the scaling magnitude used in the third experiment where a larger $R_{0}^{true}$ was chosen to more clearly demonstrate the potential for estimation bias. 

\subsubsection{Effect of the initial viral load of the index case on decoherence}
To understand the relationship between $V(0)_{index}$ and the convergence of transmission dynamics in an outbreak, we varied $V(0)_{index}$ and measured the effect on the rate of decoherence in the distribution of initial viral loads (see Methods section \ref{decoher}). We investigated five initial viral load values---$V(0)_{index}=4.5,14.2,45,142$ \& $450$---and initialised 200 outbreaks under each setting. We simulated each outbreak until 10,000 cases were recorded and disregard outbreaks which did not meet this final outbreak size in order to ensure a well-sampled distribution of initial viral loads. We measured the rate of decoherence using the described method (see Section \ref{decoher}). We pooled the initial viral load data for each generation across simulations to capture the typical behaviour of the system. Similarly to the first experiment, we scaled $P_{trans}$ (see Equation \ref{ptrans}) such that $R_{0}^{true} \approx 2$ (see Supplementary Material \ref{sup3}).

\subsubsection{Effect of the initial viral load of the index case on estimation of transmission properties}
In this experiment, we aimed to establish how any link between $V(0)_{index}$ and early outbreak dynamics determined in the analysis of the decoherence process (Section \ref{decoher}), translated into a bias in the estimation of transmission properties. We used two main parameter sets in this experiment. First, we investigated three $V(0)_{index}$ values---$V(0)_{index}=4.5,45$ \& $450$---and scaled $P_{trans}$ (see Equation \ref{ptrans}) such that the $R_{0}^{true} \approx 2$ (see Supplementary Material \ref{sup3}). Second, we investigated a low $V(0)_{index}$ value---$V(0)_{index}=4.5$---and scaled $P_{trans}$ (see Equation \ref{ptrans}) such that the $R_{0}^{true} \approx 4$ (see Supplementary Material \ref{sup3}).
\\\\
We initialised 200 outbreaks for each $V(0)_{index}$ value, and simulated each up to 50 days. We disregarded simulations in which the outbreak dies out (i.e. no infected individuals) before 50 days. We applied the transmission property estimation methods described (Section \ref{est}) each day between a range of 18--50 days of each outbreak. We did not analyse the first 17 days, as the extremely limited case data meant the methods were unlikely to be reliable \cite{cori2013new}. At each observation point, the data consists only of the serial intervals (corrected for right truncation) and symptom onset incidence observed up to that day. Each observation was designed to simulate how estimation would proceed when no previous sources of data are available (i.e. observing a novel pathogen).
\\\\
We assumed model conditions that allowed us to assess $R_0$. In particular, we assumed a large, well-mixed population that ensured susceptible depletion is not substantial early in an outbreak, and we assume host behaviour is unchanged over the observed time frame. These assumptions mean the transmission dynamics were governed by the same set of parameters over the whole simulation. As such, we set the lower bound of the time window for computing the reproduction number ($t - \tau$) to zero. This setting meant we analysed the entire outbreak observed up to time $t$ when estimating $R_0$ and the time window width ($\tau$) would grow as $t$ increased.

\section{Results}
\subsection{Effect of the initial viral load of the index case on outbreak growth}

\begin{figure}[p!]
    \centering
    \includegraphics[scale=0.5]{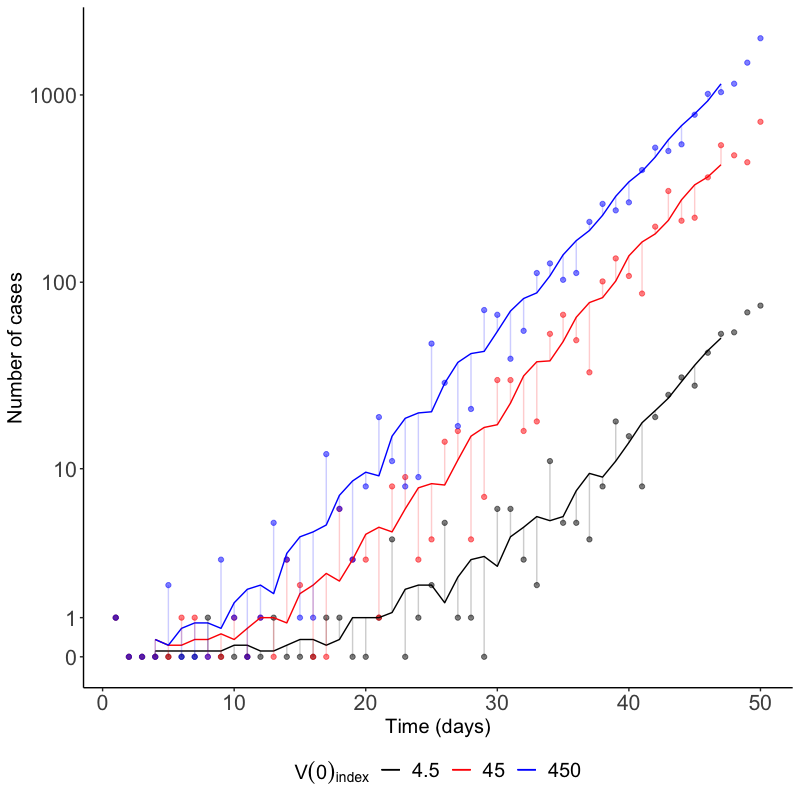}
    \caption{Time series plots of early disease incidence of outbreaks with different index case viral loads. The early rate at which an outbreak grows increases as the initial viral load of the index case ($V(0)_{index}$) increases. The effect of the initial viral load of the index case on outbreak growth reflects a multi-generational correlation of infection dynamics between cases in a transmission chain, introduced by the transmission mechanism described. Representative incidence trajectories shown here were chosen from a set (N=200) based on a defined set of outbreak summary statistics criteria (see Supplementary Material \ref{sup6}). The 7-day smoothed mean (solid line) is shown alongside the raw data (points).}
    \label{fig:fig2}
\end{figure}

Increasing the initial viral load of the index case ($V(0)_{index}$) has a substantial effect on the shape of the resulting outbreak in the first 50 days (Figure \ref{fig:fig2}). Under low $V(0)_{index}$ settings, fewer cases emerge across the observed period and outbreaks typically grow at a slower rate. This result reflects the different index case infection dynamics between settings, and their effect on subsequent cases produced in the early outbreak generations. 
\\\\
We also observe a distinct periodicity to the growth of case incidence which produces a `sawtooth' pattern in the data around the 7-day smoothed mean (Figure \ref{fig:fig2}). This pattern emerges from correlation in the timing of the peak viral load ($t_{Vmax}$), the most infectious period for a case, between infections in transmission chains. This correlation introduces a periodic rise in new infections.

\subsection{Effect of the initial viral load of the index case on decoherence}

\begin{figure}[p!]
    \centering
    \includegraphics[scale=0.34]{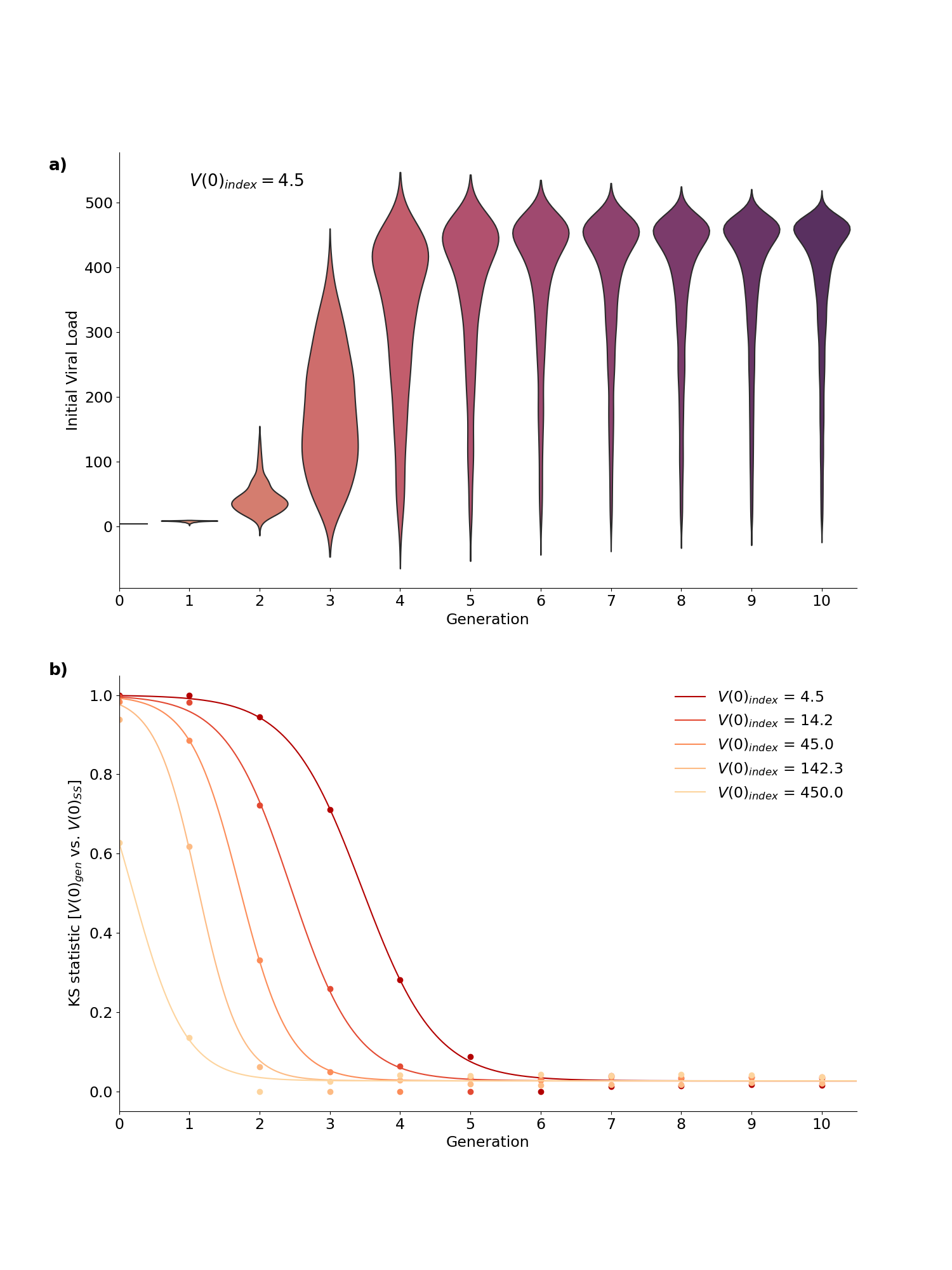}
    \caption{Dynamics of the distribution of initial viral loads as a function of contagion generation. The violin plots in subfigure (a) demonstrate how the distribution of initial viral loads over all infected cases evolves as the contagion spreads when the initial viral load of the index case is low ($V(0)_{index} = 4.5$). The trajectories of KS statistics in subfigure (b) show how the viral load distribution at each generation [$V(0)_{gen}$] approaches a converged state [$V(0)_{SS}$], occurring at some converged generation $g_c$, with the rate of approach increasing with the initial viral load of the index case. Dots represent specific KS statistic values for each generation; solid lines represent fitted logistic curves, used for illustrative purposes.}
    \label{fig:fig3}
\end{figure}

The rate at which the system converges is positively related to $V(0)_{index}$ (Figure \ref{fig:fig3}b). That is, the number of generations required before system dynamics stabilise ($g_c$), decreases as $V(0)_{index}$ increases. This trend is most clear in the lowest $V(0)_{index}$ setting simulated ($V(0)_{index}=4.5$) where 5--6 generations of outbreak growth are required before the initial viral load distribution converges (Figure \ref{fig:fig3}a). $g_c$ is substantially smaller when the highest $V(0)_{index}$ is simulated ($V(0)_{index}=450$), where convergence occurs in 2--3 generations (see Supplementary Material \ref{sup7}).
\\\\
Importantly, the relationship between $V(0)_{index}$ and the decoherence dynamics demonstrates that incidence data produced in the early generations of outbreaks may not be representative of future case dynamics. Furthermore, how representative these early cases are of the later dynamics of the system is related to the infection dynamics of the index case. We investigate how this result relates to estimations of the serial interval and $R_0$ in the following section.

\subsection{Effect of the initial viral load of the index case on estimation of transmission properties}
\subsubsection{Estimating the serial interval}

The mean serial interval estimate converges to a value of 4.2--4.3 days across the 50 days of observation, regardless of $V(0)_{index}$ (see Supplementary Material \ref{sup8}). There is typically an overestimation of the serial interval early in the outbreak observation window in the low $V(0)_{index}$ scenarios (e.g. $V(0)_{index}=4.5$). This overestimation is reflective of the longer incubation periods in infections with low initial viral loads (see Supplementary Material \ref{sup2}), which dominate these outbreaks in the early phase.

\subsubsection{Estimating $R_0$}

\begin{figure}[p!]
    \centering
    \includegraphics[scale=0.55]{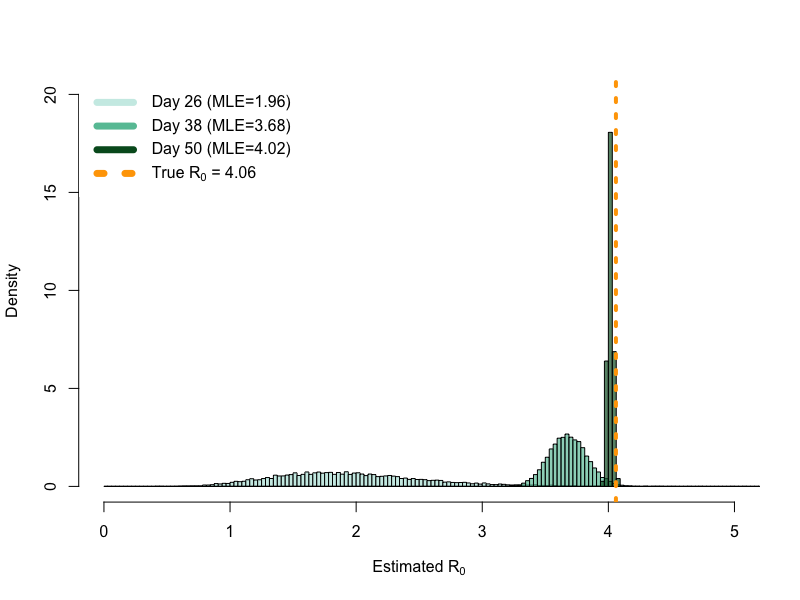}
    \caption{$R_0$ posterior distribution at three time points (days 26, 38 \& 50) for a single representative outbreak with a low $V(0)_{index}$ ($V(0)_{index}=4.5$). The central estimate of the posterior distribution increases over the course of the 50 day observation period of the outbreak, moving from 1.96 at day 26 up to 4.02 on day 50. As the contagion progresses, the estimate approaches the approximation of the true $R_0$ ($R_{0}^{true}$) of the system 4.06 (orange dashed line). The variance of the posterior decreases over the course of the outbreak, reflecting the increasing amount of case data becoming available and the increasing homogeneity of the case infection dynamics. The outbreak producing the data for this figure was selected as representative from a set (N=200) based on a defined set of outbreak summary statistics criteria (see Supplemental Material \ref{sup6}).}
    \label{fig:fig4}
\end{figure}

\begin{figure}[p!]
    \centering
    \includegraphics[scale=0.55]{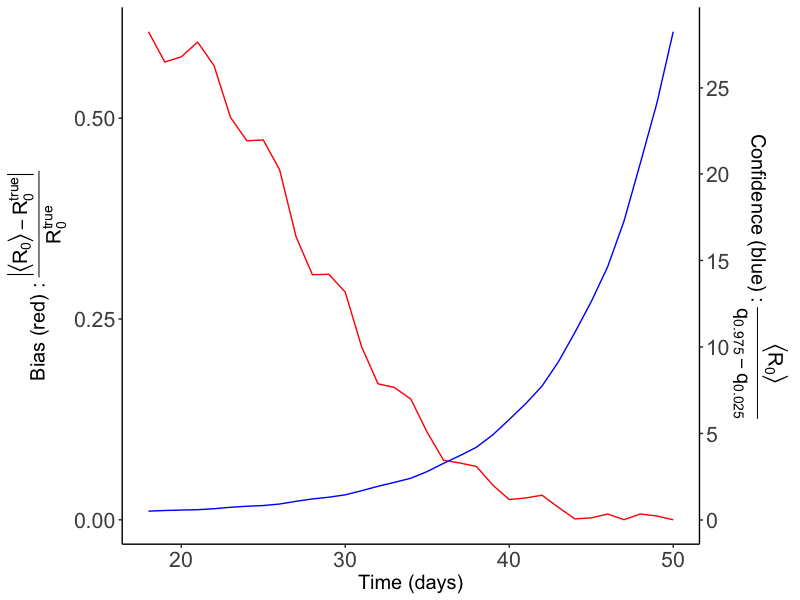}
    \caption{Relationship between relative bias and confidence of reproduction number estimation over time. The bias (measured as the relative difference between the estimated and the true value of $R_0$) decreases toward zero, as the initial conditions approach the relaxed state. As bias decreases, confidence increases (measured as the ratio of the estimated value and the 95\% inter-quantile range), as the number of observed cases and similarity of the new cases increases.}
    \label{fig:fig5}
\end{figure}

$\langle R_{0} \rangle_{50}$ converges to a value of 2--2.1 across the 50 days of observation, regardless of the $V(0)_{index}$ (see Supplementary Material \ref{sup9}). In the early period of observation (days 18--30), the central estimate ($\langle R_{0} \rangle_{d}$, where $d$ is between 18--30) of the lower initial viral load scenarios is typically an underestimate of the eventual converged value; for example, the lowest $V(0)_{index}$ setting ($V(0)_{index}=4.5$) produces  $\langle R_{0} \rangle_{d}$ across days ($d$) 18--30 of between 1.6--1.8. The underestimation of $R_0$ in these scenarios can be attributed to the fact that cases with lower initial viral loads will typically transmit to fewer individuals due to their lower peak viral load. As these cases typically dominate the early period of the low initial viral load scenarios, lower transmission is observed.
\\\\
We also examined $R_0$ estimation bias for a low $V(0)_{index}$ setting ($V(0)_{index}=4.5$) when $R_{0}^{true}$ is higher---$R_{0}^{true} \approx$ 4. Simulating a larger $R_{0}^{true}$ alongside our previous analysis allowed us to understand how the bias persists in a more infectious outbreak scenario. We increased $R_{0}^{true}$ by applying a scaling factor of 2 to our definition of $P_{trans}$ (see Supplementary Material \ref{sup3}).
\\\\
Comparing the effect of a higher transmission rate on the development of the outbreak, the magnitude of the early bias increases, but the period of time for convergence to occur remains similar (see Supplementary Material \ref{sup10}). Examining a representative outbreak with $R_{0}^{true} \approx$ 4, we observe a substantial underestimation of $R_0$ over the first 40 days of the outbreak (Figure \ref{fig:fig4}). The uncertainty in these estimates is initially high (day 26---95\% IQR: 0.968--2.389) and decreases across the observation period (day 38---95\% IQR: 3.389--3.983 \& day 50---95\% IQR: 3.977--4.057).
\\\\
We computed the mean bias and confidence (see Equations \ref{mean_bias} \& \ref{mean_conf}) from the set of outbreak simulations at each day across our observation window (Figure \ref{fig:fig5}). $Bias'$ decreases over the course of the observation period, falling from $Bias'$ between 0.3--0.6 across days 18--30 to $Bias'\approx0$ across days 40--50. This decrease reflects the convergence of the case dynamics that occurs across this period. $Confidence'$ increases over the course of the observation period from $Confidence'\approx0$ across days 18--25 to $Confidence'$ between 5--25 across days 40--50. This rise in $Confidence'$ is reflective of the increasing case count and the increasing homogeneity in the infection dynamics of newly infected individuals. Figure \ref{fig:fig4} demonstrates the bias magnitude and confidence in the reproduction number estimate for a single outbreak as a function of time.

\section{Discussion}
Understanding the transmissibility of a pathogen is critical for designing effective policies to reduce its spread in a community. Accurate estimation of transmission properties, such as $R_0$, is crucial for establishing the scale of risk posed in future outbreaks. Producing accurate estimates of transmission properties with limited observed data is difficult, particularly when there is substantial heterogeneity in the disease characteristics of the infected population.
\\\\
In this study, we demonstrated that assumptions made about viral transmission can lead to biased estimates of the serial interval and $R_0$. We used a novel decoherence analysis method, together with previously published serial interval and reproduction number estimation procedures, to demonstrate that early outbreak dynamics are substantially dependent on the initial viral load of the index case, $V(0)_{index}$ (Figures \ref{fig:fig2} \& \ref{fig:fig3}), and hence estimates of transmission properties based on observations in this period could be biased (Figures \ref{fig:fig4} \& \ref{fig:fig5}).

\subsection{The initial viral load of the index case affects the rate at which the system converges to a steady state}
To simulate disease spread, we introduced a multiscale model describing transmission of a virus in a well-mixed, susceptible population. We described a within-host model with viral load dynamics dependent on the size of an initial viral load. Infected individuals had their transmission potential scaled by their viral load via a sigmoidal relationship. Additionally, the initial viral load of an infectee was scaled by the viral load of their infector at transmission, also via a sigmoid function. The influence of the index case(s) infection dynamics on outbreak development has been explored in similar simulation studies \cite{mercer2011effective, steinmeyer2010methods}. Using our model of disease transmission, we demonstrated $V(0)_{index}$ affects the rate at which a population-level decoherence process occurs, during which the distributions of initial viral loads in each subsequent generation converge to a steady state (Figure \ref{fig:fig3}).
\\\\
The shape of the distribution of initial viral loads during the decoherence process reflects key aspects of the underlying model of viral transmission (Figure \ref{fig:fig3}). The emergence of the positively skewed distribution of initial viral loads in the infected population reflects the tendency for higher initial viral load cases to be more infectious, due to their increased peak viral loads. Once cases with high initial viral loads (i.e. $V(0)>400$) appear during an outbreak, they will typically begin to dominate the infected population (see generations 2--5 in Figure \ref{fig:fig3}a). This process is reflected in the logistic shape of the decoherence trajectories of the low $V(0)_{index}$ scenarios (see $V(0)_{index}=4.5,14.2,45.0$ in Figure \ref{fig:fig3}b). In these scenarios, the distribution of initial viral loads in early generations are substantially dissimilar to the converged generation $g_c$. The generational distribution then transitions to the converged distribution once high initial viral load cases appear and begin to dominate.

\subsection{The initial viral load of the index case affects the accuracy of transmission parameter estimates}
Having identified that $V(0)_{index}$ could exert a substantial influence on the development of an outbreak, we then investigated its effect on estimation of transmission properties. We estimated the serial interval distribution and $R_0$ across the first 50 days of outbreaks seeded with index cases with varying initial viral loads. We observed that when outbreaks are seeded with low initial viral load cases, the serial interval is typically overestimated and $R_0$ is typically underestimated over the course of the early phase of the outbreak.
This result has a clear correspondence with our analysis of the outbreak decoherence process for lower $V(0)_{index}$ scenarios. Specifically, our finding that there would likely be an extended period where case infection dynamics were not representative of later dynamics, is clearly reflected in the tendency to overestimate the serial interval and underestimate $R_0$ early in an outbreak.
\\\\
Other analyses of outbreak data have found that if early transmission of a disease occurs in a subpopulation with a higher or lower intrinsic transmission rate, then the early estimates for the reproduction number will be biased \cite{mercer2011effective}. In these circumstances, the subpopulation with a higher or lower intrinsic transmission rate is over-represented in the early observations of an outbreak. For example, an outbreak of influenza (which is typically more transmissible among children \cite{opatowski2011transmission}) seeded in a school environment, where younger individuals make up a high proportion of the population. In this study, we demonstrated an alternate way that cases with a lower intrinsic transmission rate could emerge early in an outbreak, based solely on the mechanistic features of transmission in our multiscale model, and how such cases could then lead to biased transmission property estimation. Specifically, as opposed to subpopulations defined on host characteristics such as age or gender, we observed the initial viral load acting as an indicator of transmission potential and correlating between generations. When outbreaks are seeded by lower initial viral load cases, we observed an early over-representation of cases with low initial viral loads which biased transmission property estimates. 
\\\\
In this work, we demonstrated the potential for reproduction number estimates to be biased down due to a transmission process in which infected individuals produce strong correlations in transmission dynamics between generations. We note that this particular result (a strong, downward bias) depends on the choices we made in crafting our model of host-to-host transmission. For example, we chose to assume that higher viral loads produce a higher probability of transmission. While this is a biologically reasonable assumption \cite{handel2015crossing}, it ignores the possibility that behavioural responses could limit contact patterns due to the expression of severe symptoms associated with high viral loads (endogenous behaviour). An alternative model could suppose that low viral loads are in fact more likely to produce transmission, in which case the converged distribution of initial viral loads would be skewed to lower values. In such a case, the bias could be inverted, that is, if the index case has a viral load that is much higher than that of the average individual in the converged state, we would expect the initial reproduction number estimates to be biased upward. We suppose that the nature of estimation bias produced by correlations in viral load trajectories between generations will depend strongly on both the manifestation of disease, and the endogenous behavioural response of infected (and susceptible) individuals.

\subsection{Estimation bias could be operationally significant}
We investigated the operational significance of this bias by analysing the relative bias and confidence associated with estimates of the reproduction number across the 50-day observation period (Figures \ref{fig:fig4} \& \ref{fig:fig5}). We observed a central phase of the observation window (between days 30--40) where there was a potential for a biased estimate to be produced with a substantial degree of confidence. Such underestimation, in this case, of the reproduction number could have a serious impact on the perception of a risk in a community, and associated public health decisions. To highlight the potential for these estimates to be problematic, we assessed how the estimate could be used to forecast future outbreaks of the same pathogen in similar communities. Using a standard SIR compartmental model without vital dynamics, we adjusted the recovery rate and transmission rate parameters to compare how modelling would differ using the central $R_0$ estimates at day 26 ($R_0$ = 1.96), day 38 ($R_0$ = 3.68) and day 50 ($R_0$ = 4.02) of the representative outbreak sample shown in Figure \ref{fig:fig4} (see Supplementary Material \ref{sup11}). Compared to the earlier day 26 estimate, using the Day 38 and day 50 estimates resulted in an increase of approximately 155\% and 176\% in the peak number of infected individuals and a decrease of 58.3\% and 62.5\% in the time taken to reach the peak, respectively. The size and timing of the peak number of infected individuals are critical measures in planning public health responses and the difference shown here demonstrates the importance of the underestimation for policy decision makers.
\\\\
Furthermore, media coverage of an emerging pathogen, as has been witnessed in the current COVID-19 pandemic, has a profound effect on the perception of a disease in a population \cite{diaz2022stock,liu2021role,kubiczek2021challenges}. Biased estimates, when reported to the public, have the potential to induce a sense of panic (over-estimation) or complacency (under-estimation) in public perception of risk due to a pathogen. Media coverage plays a role, alongside public policy formulation, in determining the public response to an outbreak which can have substantial effects on disease spread. The perception of a pathogen can also have serious effects on the economic output and mental wellbeing of a community \cite{diaz2022stock}.
\\\\

\subsection{Limitations and future work}
Some key assumptions of the underlying outbreak model should be kept in mind when interpreting the results of our study. The determination of initial viral load in an infected individual and its resulting effect on disease progression are by no means established relationships in the study of viruses. Experimental data around the initial viral load are difficult to obtain, mainly due to the difficulty of measuring early infection dynamics and the establishing dose in infected hosts. Similarly, the relationship between host viral load and infectiousness is another relationship that is not well characterised \cite{handel2015crossing}. Most studies assume some positive relationship between the two quantities, commonly taking the form of a linear, logarithmic or sigmoidal form \cite{handel2015crossing,hart2020theoretical,handel2013multi}. A possible avenue for future work would be to investigate estimation bias emerging from alternate representations of transmission.
\\\\
Similarly, the assumption of perfect observation in the disease surveillance of the population should be reflected upon when interpreting these results. While the nature of the reproduction number estimation method means if we have a consistent proportion of unobserved cases, we should still obtain accurate estimates, the likelihood of this proportion remaining steady across an outbreak is low \cite{thompson2019improved}. Furthermore, the heterogeneous viral loads across the infected cases in our model may impact the likelihood of detecting some cases. Specifically, if severity of symptoms was positively related to the viral load of a case and the likelihood of observation was related to the presence of symptoms, it may mean lower viral load cases are less likely to be detected. An interesting extension of our model could be to see how relating likelihood of detection to viral load affects the quality of the estimates produced from the discussed estimation procedures.
\\\\
Another avenue for further work is to investigate how the bias explored in this analysis would play out in a later phase of pathogen emergence, where multiple outbreaks have been observed in different settings. One could explore how the composition of the initial conditions describing a set of simulated observed outbreaks affects the aggregate transmission property estimates. The challenge in this scenario is effectively incorporating data from different outbreaks together to produce one descriptive estimate.  Hierarchical models have been used effectively to estimate disease parameters across different environments \cite{alahakoon2022estimation}.

\subsection{Conclusion}
In summary, we developed a stochastic model of disease transmission, accounting for the transfer of virus from infected to susceptible hosts through the use of multiscale modelling. We found the index case initial conditions were influential in the development of an outbreak. Finally, we demonstrated the potential for a bias to exist in both the estimation of the serial interval and $R_0$ in the first 50 days of an outbreak. This investigation illustrates how multiscale models---that explicitly capture the relationship between host-level viral dynamics and population-level transmission dynamics---can be used to evaluate widely-used methods for estimating transmission properties.

\subsection{Acknowledgements}
I would like to acknowledge the contribution of Dr. David Price, {\it The Peter Doherty Institute for Infection and Immunity}, to the development of the methodology of this work. All simulations were performed using the Nectar Research Cloud (project Infectious Diseases), a collaborative Australian research platform supported by the National Collaborative Research Infrastructure Strategy (NCRIS).

\subsection{Data accessibility}
All analyses were performed in python3 \& R. Data and code is available at \url{https://doi.org/10.5281/zenodo.7297085}.

\bibliographystyle{vancouver}
\newpage
\section{References}
\bibliography{references2}

\newpage

\newcommand{\beginsupplement}{%

 \setcounter{table}{0}
   \renewcommand{\thetable}{S\arabic{table}}%
   
     \setcounter{figure}{0}
      \renewcommand{\thefigure}{S\arabic{figure}}%
      
      \setcounter{page}{1}
      \renewcommand{\thepage}{S\arabic{page}} 
      
      \setcounter{section}{0}
      \renewcommand{\thesection}{S\arabic{section}}
      
      \setcounter{equation}{0}
      \renewcommand{\theequation}{S\arabic{equation}}
     }


\beginsupplement

\FloatBarrier
{\bf \Large{Correlation of viral loads in disease transmission chains could bias early estimates of the reproduction number: Supplementary Material}}
\section{Within-host parameter values} \label{sup1}
The parameter values outlined in Table \ref{tab:WH_params} describe the within-host model across all simulations in this study. The initial viral load ($V(0)$) is not described here; the method for determining this value based on the donor viral load at transmission, is described in Supplementary Material \ref{sup4}. Some parameters related to the strength of the immune response are sampled from Gamma distributions, with a standard deviation relative to the size of the mean.
\\\\
The viral load threshold at which infected individuals with a greater viral load will be defined as `infectious' ($V_T$) is set to $V_T = 13431.67$. This value was chosen as it aligns with the point at which the initial viral load in any recipient of infection from this host will be $1$ (i.e. $V_R(0)=1$). This means all new infections will be initiated with a viral quantity greater than $1$. The value for $V_T$ can be computed using Equation \ref{eq:f_sig} in Supplementary Material \ref{sup4}, and solving for $V_d$ when $V_R(0)=1$.
\\\\
The index case will have the same within-host parameter values, with two key differences. The initial viral load ($V(0)$) will be defined at the initialisation of the simulation based on the experiment being conducted. And, the mean values for the parameters sampled from a Gamma distribution will be used to describe these parameters, to ensure consistency in the index case across simulations.
\label{sup1}
\begin{table}[]
\centering
\caption{Parameter values for within-host mathematical model.}
\begin{tabular}{|l|l|}
\hline
\textbf{Parameter}     & \textbf{Value}                                   \\ \hline
$r$                      & 5.5                                              \\ \hline
$k_I$                   & Gamma($Mean=0.05$, $SD=0.05 \times 0.005$)                                             \\ \hline
$k_N$                   & Gamma($Mean=0.5$, $SD=0.5 \times 0.005$)                                              \\ \hline
$k_P$                   & Gamma($Mean=2$, $SD=2 \times 0.005$)                                                \\ \hline
$a_I$                   & $10^{-9}$                       \\ \hline
$a_N$                   & $10^{-8}$                      \\ \hline
$a_P$                   & $5\times 10^{-6}$ \\ \hline
$b_I$                   & 2                                                \\ \hline
$d_N$                   & 0.05                                             \\ \hline
$c$                      & 0.01                                             \\ \hline
$K_V$                   & Gamma($Mean=10^{11}$, $SD=10^{11} \times 0.005$)                      \\ \hline
$K_I$                   & Gamma($Mean=100$, $SD=100 \times 0.005$)                                              \\ \hline
$\tau_N$                & 2.5                                              \\ \hline
$\tau_P$                & 3                                                \\ \hline
$N(0)$                  & 0                                                \\ \hline
$P(0)$                  & 2                                                \\ \hline
\end{tabular}
\label{tab:WH_params}
\end{table}

\begin{figure}[p!]
    \centering
    \includegraphics[scale=0.60]{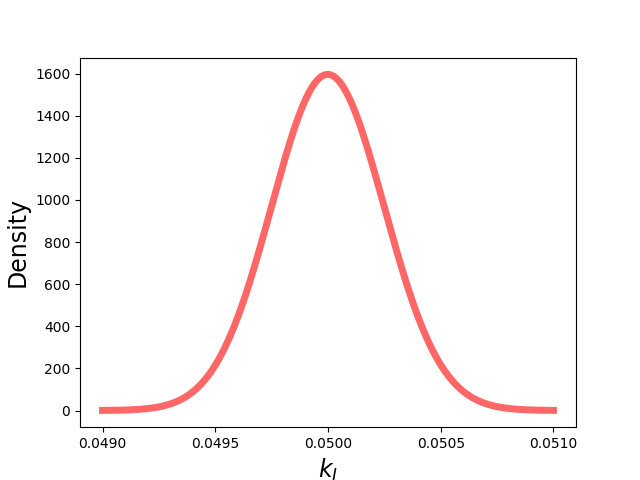}
    \caption{The probability density function of the gamma distribution in which the value for the rate of viral removal relevant to the innate immune response ($k_I$) is sampled for each newly infected individual.}
    \label{fig:inc_period}
\end{figure}

\begin{figure}[p!]
    \centering
    \includegraphics[scale=0.60]{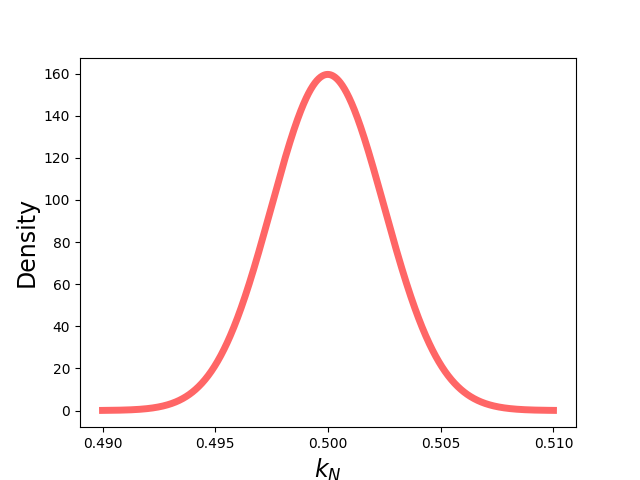}
    \caption{The probability density function of the gamma distribution in which the value for the rate of viral removal relevant to the non-specific memory cells ($k_N$) is sampled for each newly infected individual.}
    \label{fig:inc_period}
\end{figure}

\begin{figure}[p!]
    \centering
    \includegraphics[scale=0.60]{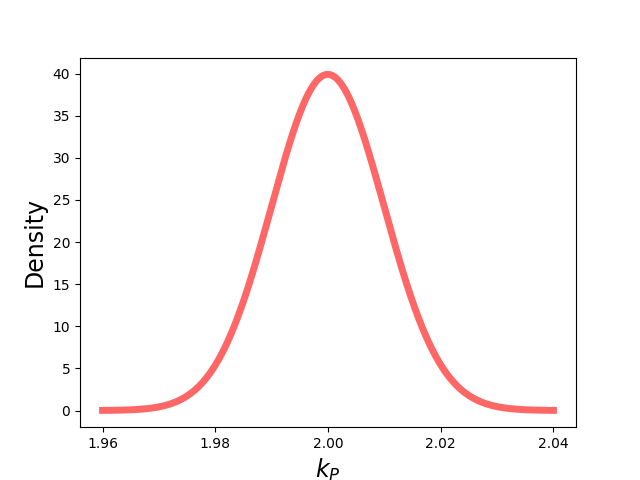}
    \caption{The probability density function of the gamma distribution in which the value for the rate of viral removal relevant to the specific memory cells ($k_P$) is sampled for each newly infected individual.}
    \label{fig:inc_period}
\end{figure}

\begin{figure}[p!]
    \centering
    \includegraphics[scale=0.60]{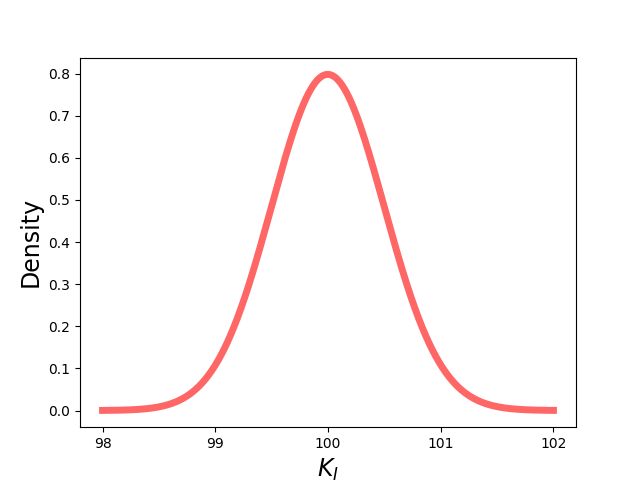}
    \caption{The probability density function of the gamma distribution in which the value for the innate immunity carrying capacity ($K_I$) is sampled for each newly infected individual.}
    \label{fig:inc_period}
\end{figure}

\begin{figure}[p!]
    \centering
    \includegraphics[scale=0.60]{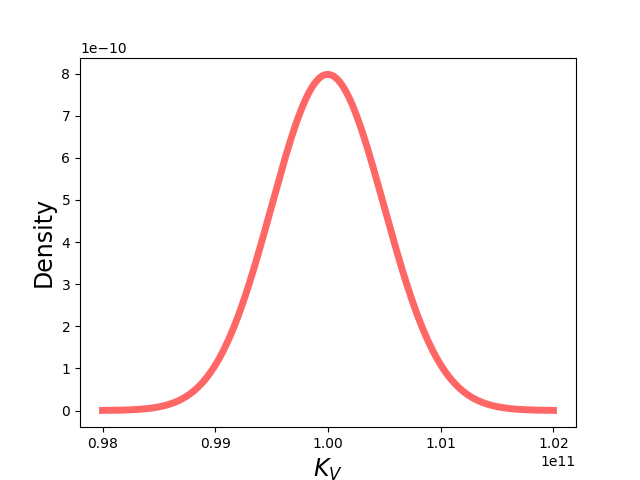}
    \caption{The probability density function of the gamma distribution in which the value for the viral load carrying capacity ($K_V$) is sampled for each newly infected individual.}
    \label{fig:inc_period}
\end{figure}

\section{Within-host viral load response curves}
\label{sup2}
\begin{figure}[p!]
    \centering
    \includegraphics[scale=0.28]{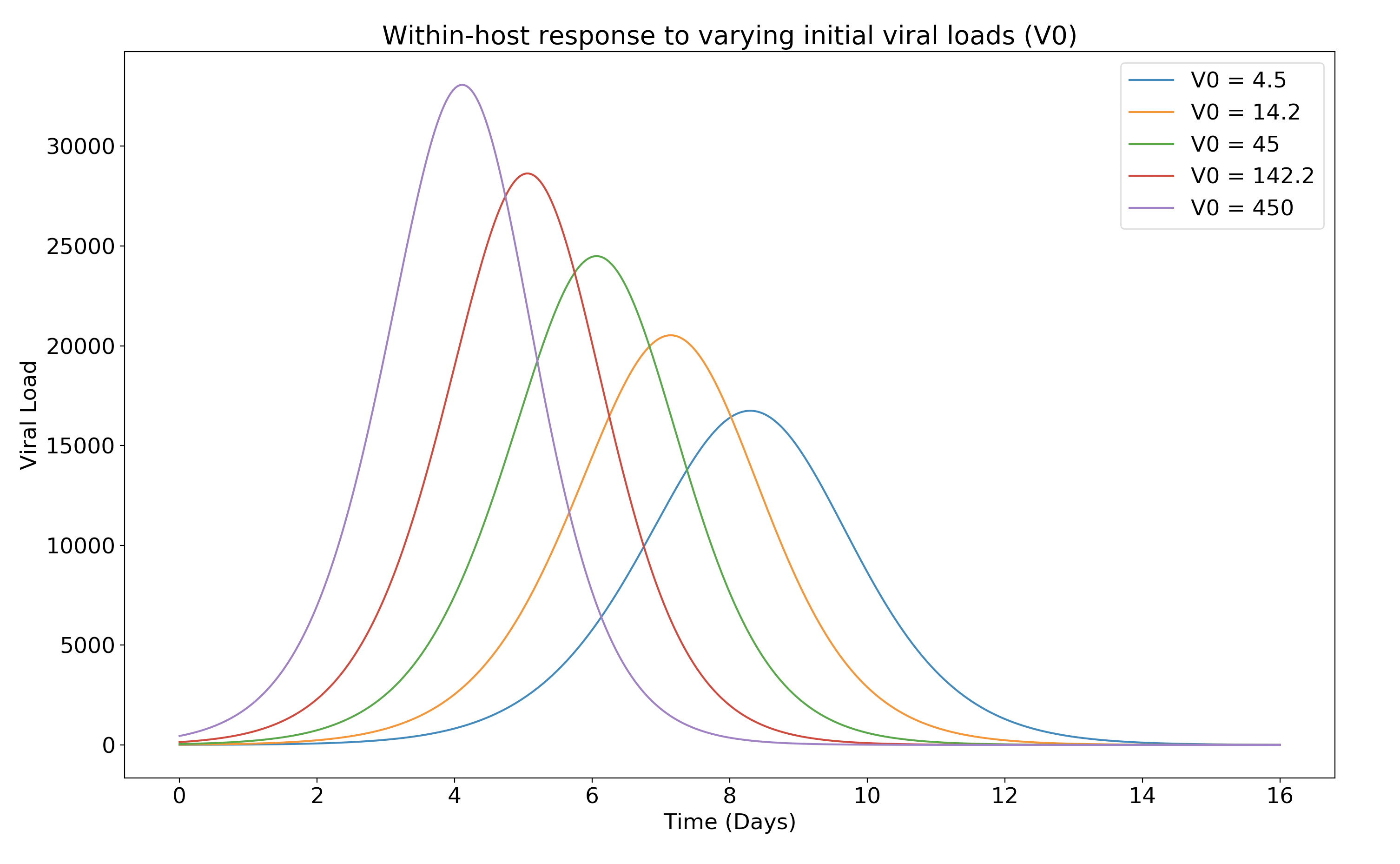}
    \caption[Within-host viral response to varying initial viral load]{The within-host model viral load curve is affected by the initial viral load. As the initial viral load increases, viral load peak increases, time to peak decreases and infection duration decreases. These effects outline the significance of the initial viral load in the viral load dynamics of this within-host model and inform analysis of spread dynamics when implemented in a multiscale model. `V0' is used in the legend to refer to the initial viral load value.}
    \label{fig:curvesv0}
\end{figure}

The initial viral load has a substantial effect on the shape of the viral load curve (Figures \ref{fig:curvesv0} \& \ref{fig:inc_period}). As the initial viral load increases, viral load peak increases, time to peak decreases and infection duration decreases. The within-host parameters sampled from a Gamma distribution are held constant in the comparison to clearly illustrate the effect of the initial viral load on the within-host experience.

\begin{figure}[p!]
    \centering
    \includegraphics[scale=0.60]{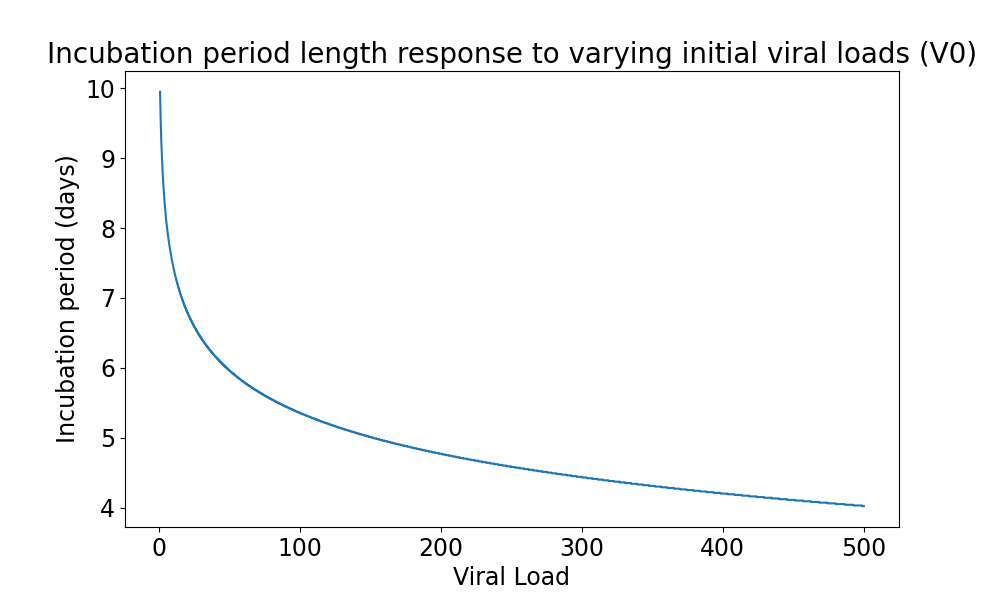}
    \caption[Within-host viral response to varying initial viral load]{The within-host model incubation period is affected by the initial viral load. As the initial viral load increases, the length of the incubation period decreases. These effects outline the significance of the initial viral load in the viral load dynamics of this within-host model and inform analysis of spread dynamics when implemented in a multiscale model. Symptom onset is approximated by the time when peak viral load is reached.}
    \label{fig:inc_period}
\end{figure}

\section{Host viral load to probability of transmission mapping}
\label{sup3}
\begin{figure}[p!]
    \centering
    \includegraphics[scale=0.48]{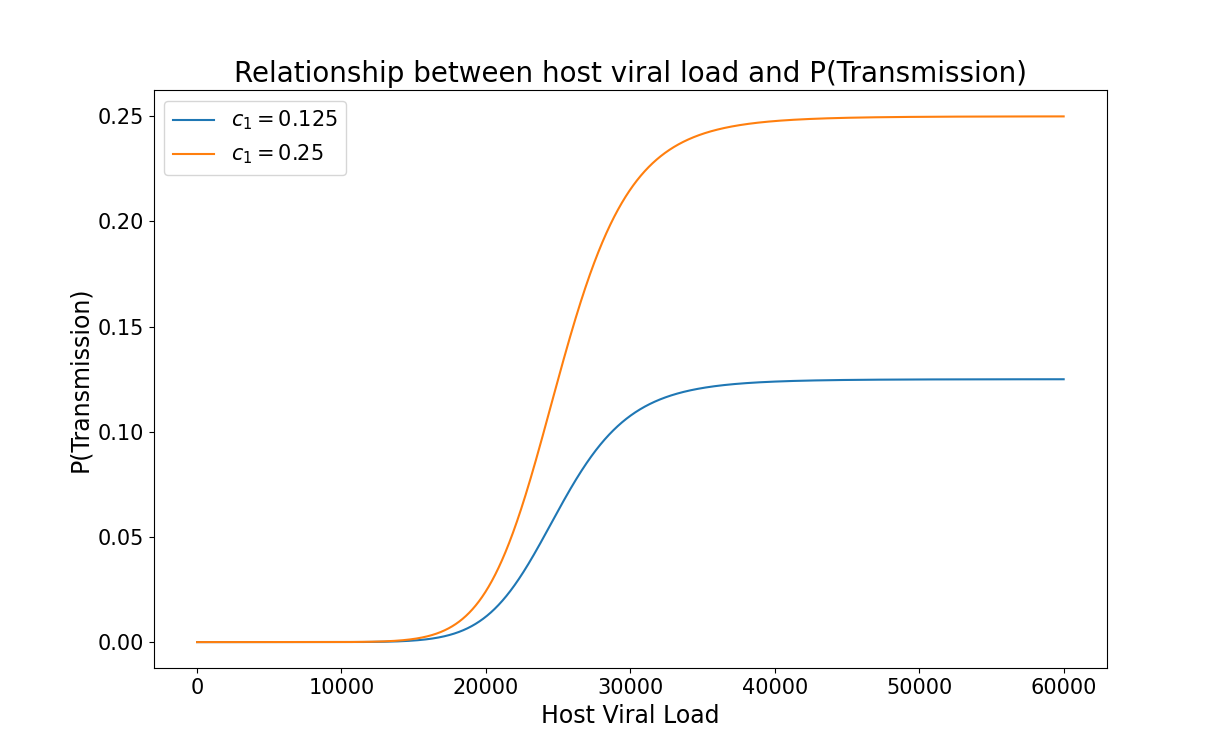}
    \caption{Sigmoidal relationship between host viral load and the probability of transmission.}
    \label{fig:trans_map}
\end{figure}

We define a sigmoidal mapping between donor viral load and the probability of transmission (Figure \ref{fig:trans_map}). The function relating donor viral load ($V_d$) to the probability of transmission given contact exposure ($P_{trans}$) is given by:
\begin{equation}
    \label{eq:g_sig}
    P_{trans} = \frac{c_1V_{d}^{\zeta}}{V_{d}^{\zeta}+c_2^{\zeta}} \,,
\end{equation}
where $c_2=25000$ and $\zeta=10$. For experiment 1 and 2, $c_1=0.125$  and for experiment 3, $c_1=0.25$.

\section{Donor viral load to recipient initial viral load mapping}
\label{sup4}
\begin{figure}[p!]
    \centering
    \includegraphics[scale=0.55]{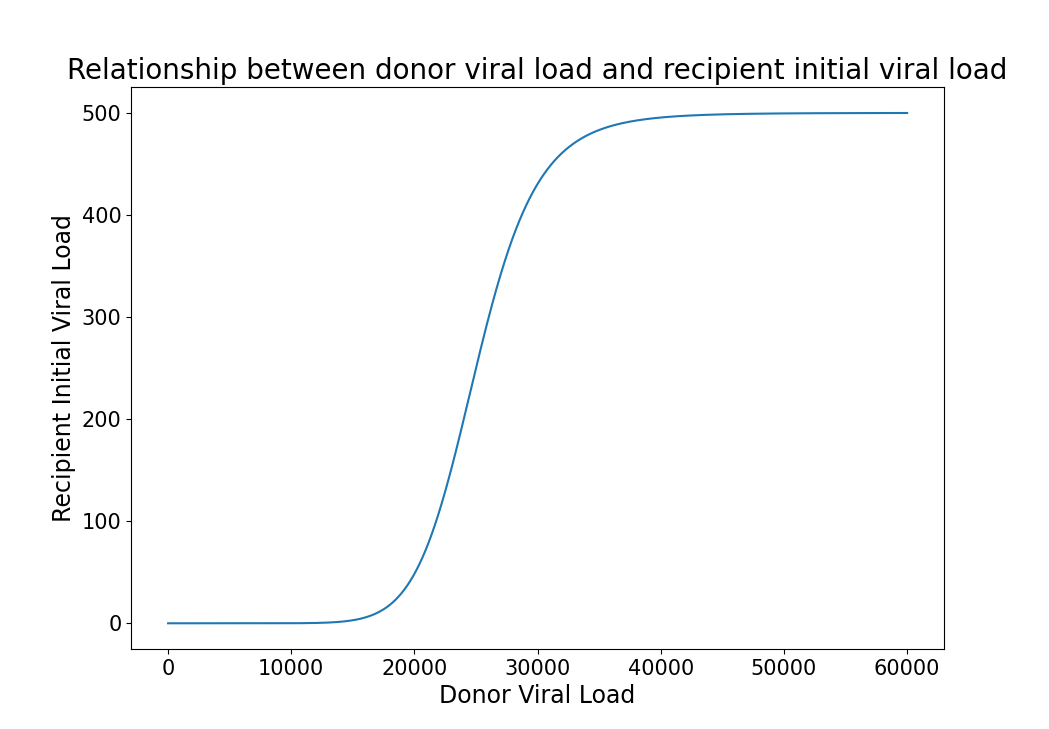}
    \caption{Sigmoidal relationship between donor viral load at transmission and the recipient initial viral load.}
    \label{fig:dv_r0_map}
\end{figure}

We define a sigmoidal mapping between donor viral load at transmission and the recipient initial viral load (Figure \ref{fig:dv_r0_map}). The function relating donor viral load ($V_d$) to the recipient initial viral load ($V_R(0)$) is given by:
\begin{equation}
    \label{eq:f_sig}
    V_R(0) = \frac{d_1V_d^{\kappa}}{V_d^{\kappa}+d_2^{\kappa}} \,,
\end{equation}
where $d_1=500$, $d_2=25000$ and $\kappa=10$. 

\section{Transmission property estimation workflow} \label{sup5}
\begin{figure}[p!]
    \centering
    \includegraphics[scale=0.5]{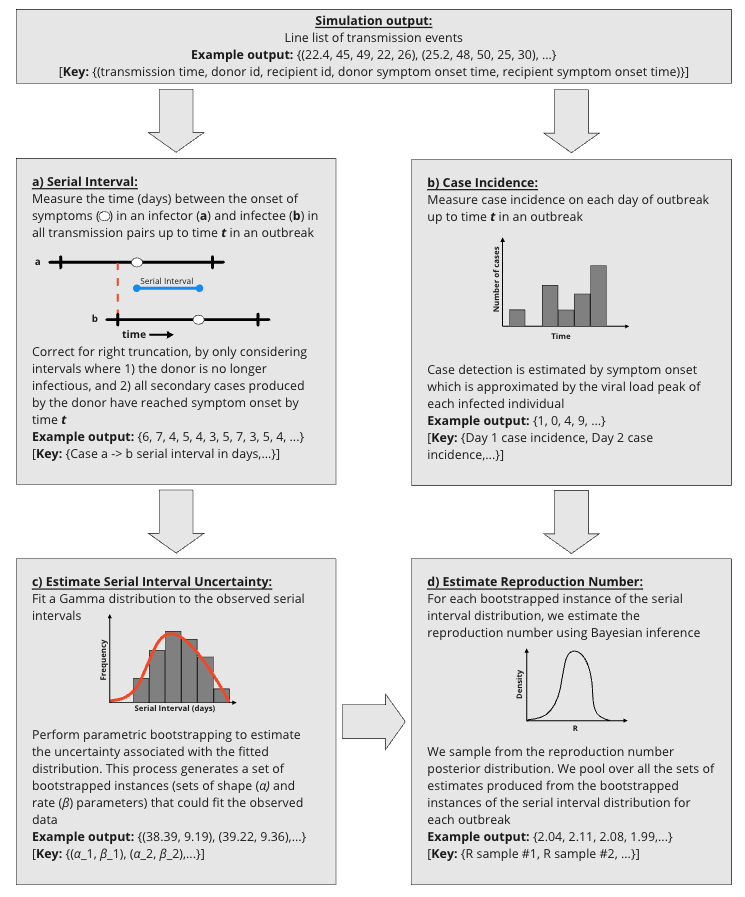}
    \caption{Transmission property estimation workflow. Here, we detail the process for estimating the serial interval and reproduction number at some time $t$ in an outbreak from the simulation output. Panel (a) describes how we extract serial interval data from the simulation output. We record a set of whole numbers describing the serial intervals in days of each transmission pair. Panel (b) describes how we extract case incidence from the simulation output. We obtain a time series of daily case incidence across an outbreak. Panel (c) describes how we estimate the serial interval uncertainty using parametric bootstrapping. We generate a set of tuples describing the values of the shape ($\alpha$) and rate ($\beta$) parameters in the fit for each bootstrap iteration. Panel (d) describes how we estimate the reproduction number from the serial interval distribution fit and the outbreak case incidence. We generate a set of $R_0$ estimates, sampled from the reproduction number posterior distribution.}
    \label{fig:est_workflow}
\end{figure}

We estimate the reproduction number for each outbreak at some day $t$ using a Bayesian inference method (Figure \ref{fig:est_workflow}). This procedure is similar to the extension of the Cori {\it et al.} method \cite{cori2013new} introduced in \cite{thompson2019improved}. The key difference being we use parametric bootstrapping to estimate the uncertainty in the serial interval as opposed to an Markov Chain Monte Carlo (MCMC) approach.
\\\\
We measure serial intervals between transmission pairs and daily case incidence up to some time $t$ in an outbreak (see panels 'Serial Interval' \& 'Case Incidence' in Figure \ref{fig:est_workflow}). Symptom onset of each case is approximated as the time at which an individual reaches their viral load peak. We assume cases are detected perfectly and immediately at symptom onset. We correct the serial interval data for right truncation to ensure there is no biasing towards smaller serial intervals which are naturally observed earlier in an infectious period. 
\\\\
We fit a Gamma distribution to the measured serial intervals and estimate the uncertainty associated with the serial interval through parametric bootstrapping (see panel `Estimate Serial Interval Uncertainty' in Figure \ref{fig:est_workflow}). We use the `fitdistrplus' library in R to fit a Gamma distribution by maximum likelihood estimation. We note here this fitting process fails in the rare scenarios when all observed serial interval values are identical; if this occurs, we introduce some minor noise ($10^{-5}$) to the observed serial intervals, to allow fitting to succeed. This adjustment does not affect the resulting estimates.
\\\\
We use the same library to perform parametric bootstrapping---a process whereby we resample from the fitted distribution and fit a new distribution in order to estimate the uncertainty in the original fit. The output from this process is a set of $N_B=100$ bootstrap samples which are made up of a Gamma shape and rate parameter.
\\\\
We estimate the reproduction number using the Bayesian method developed by Cori {\it et al.} \cite{cori2013new} (see panel `Estimate Reproduction Number' in Figure \ref{fig:est_workflow}). We estimate the reproduction number for each bootstrap sample of the serial interval distribution and the daily case incidence recorded for the outbreak. We utilise the `EpiEstim' library in R to compute the expectation of infectivity used in the Bayesian framework for inferring the reproduction number.

\section{Representative sample selection criteria} \label{sup6}
We select a single representative outbreak from a set of simulated outbreaks using four key criteria:
\begin{enumerate}
    \item Outbreak size---the total number of people infected across the 50 day simulation period
    \item Time till case threshold---the number of days until a total case threshold is met 
    \item Time till daily incidence threshold---the number of days until a daily case incidence is met
    \item Portion of first 100 recorded serial intervals less than threshold---the number of serial intervals from the first 100 recorded cases less than some threshold number of days 
\end{enumerate}
We normalise each of the above measures across the simulation set of interest. We then compare each outbreak to the mean of the simulation set across the listed criteria. From these comparisons, we determine the closest outbreak to the typical behaviour of the set. 
\\\\
The above criteria were selected to assess the scale of an outbreak (e.g. 1), the timing of its growth (e.g. 2 \& 3), and the type of cases produced (e.g. 4).

\section{Dynamics of the distribution of initial viral loads as a function of contagion generation when the initial viral load of the index case is high ($V(0)_{index} = 450$).} \label{sup7}

\begin{figure}[p!]
    \centering
    \includegraphics[scale=0.46]{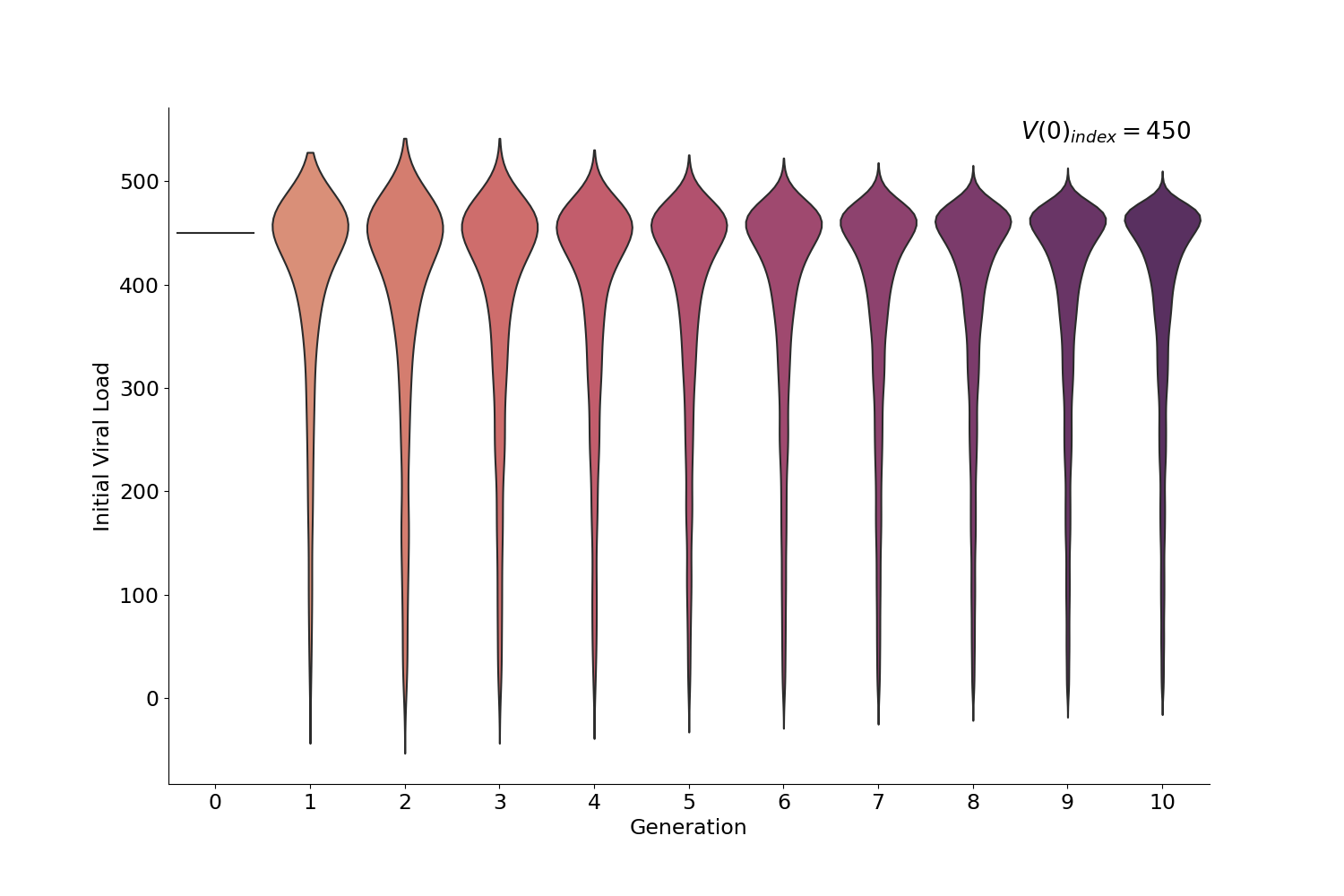}
    \caption{Dynamics of the distribution of initial viral loads as a function of contagion generation. The violin plot demonstrates how the distribution of initial viral loads over all infected cases evolves as the contagion spreads when the initial viral load of the index case is high ($V(0)_{index} = 450$).}
    \label{fig:sup7}
\end{figure}

We measured the distribution of initial viral loads as a function of contagion generation when $V(0)_{index}$ is high---$V(0)_{index} = 450$ (Figure \ref{fig:sup7}). With this seeding condition, the distribution remains consistent across case generation. Convergence occurs in the first 2--3 generations. The trend seen here contrasts substantially with the dynamics associated with outbreaks arising from index cases with low initial viral loads (see Main Text, Figure 3), where outbreaks typically take between 5--6 generations to converge.

\section{Effect of $V(0)_{index}$ on serial interval estimation} \label{sup8}

\begin{figure}[p!]
    \centering
    \includegraphics[scale=0.5]{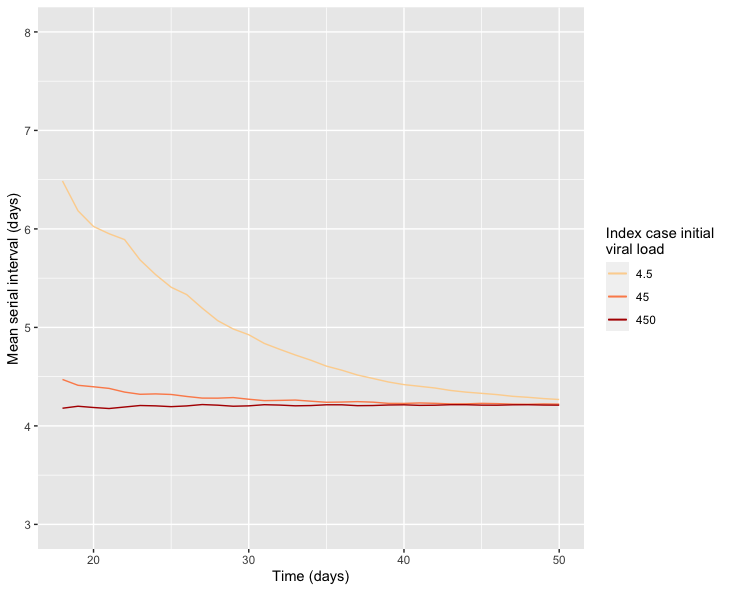}
    \caption{The mean serial interval estimate between days 18 and 50 across three different $V(0)_{index}$ sizes. Low $V(0)_{index}$ scenarios (e.g. see `4.5') result in early overestimation of the serial interval. This reflects the alternate infection dynamics of low initial viral load cases which dominate these outbreaks early. As the $V(0)_{index}$ increases, the early estimates become more reflective of later dynamics. Estimates converge to a value (approximately 4.2 days), regardless of $V(0)_{index}$ value, mirroring the decoherence observed in the previous analysis (see Figure \ref{fig:fig3}).}
    \label{fig:si_est}
\end{figure}

We estimated the serial interval each day between day 18 and 50, under different $V(0)_{index}$ settings (Figure \ref{fig:si_est}). We established the mean serial interval for a particular day and $V(0)_{index}$ by:
\begin{itemize}
    \item Fitting a gamma distribution to the observed serial interval at some day in an outbreak.
    \item Performing parametric bootstrapping to establish $N_B=100$ bootstraps of the original distribution fit
    \item Measure the mean of each parametric bootstrap, by computing:
    \begin{equation}
        mean = \frac{\alpha}{\beta} \,,
    \end{equation}
    where $\alpha$ is the shape parameter and $\beta$ is the rate parameter of the gamma distribution
    \item Pool all computed means across the simulation set of a particular $V(0)_{index}$
    \item Compute the mean of the set $N_B \times \#simulations$ of means from the bootstraps
\end{itemize}

\section{Effect of $V(0)_{index}$ on the basic reproduction number ($R_0$) estimation} \label{sup9}

\begin{figure}[p!]
    \centering
    \includegraphics[scale=0.5]{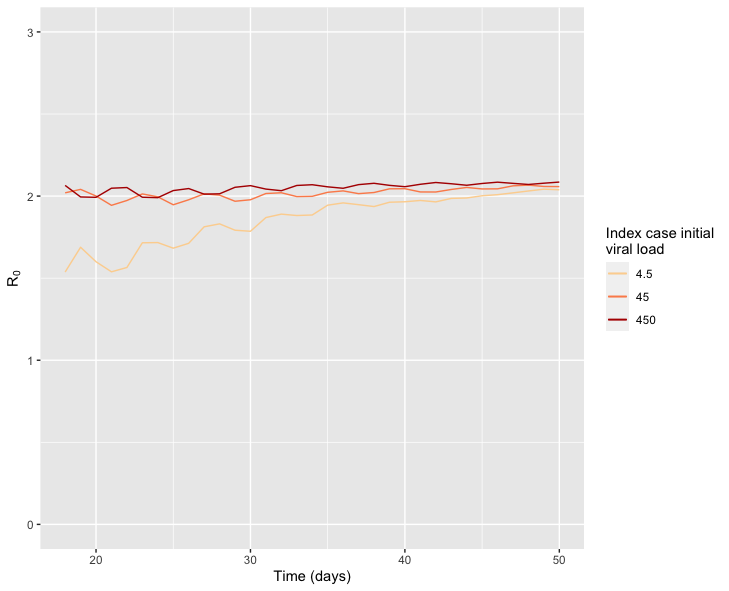}
    \caption{The mean $R_0$ estimate between days 18 and 50 across different $V(0)_{index}$ sizes. Low $V(0)_{index}$ scenarios (e.g. see `4.5') result in early underestimation of $R_0$. This reflects the alternate infection dynamics of low initial viral load cases which dominate these outbreaks early. As the $V(0)_{index}$ increases, the early estimates become more reflective of later dynamics. Estimates converge to a value (approximately 2), regardless of $V(0)_{index}$ value, mirroring the decoherence observed in the previous analysis (see Figure \ref{fig:fig3}).}
    \label{fig:r0_est}
\end{figure}

We estimated the basic reproduction number ($R_0$) each day between day 18 and 50, under different $V(0)_{index}$ settings (Figure \ref{fig:r0_est}). We established $R_0$ for a particular day and $V(0)_{index}$ through the method outlined in main text. 

\section{Effect of underlying $R_0$ on estimation bias} \label{sup10}

\begin{figure}[p!]
    \centering
    \includegraphics[scale=0.5]{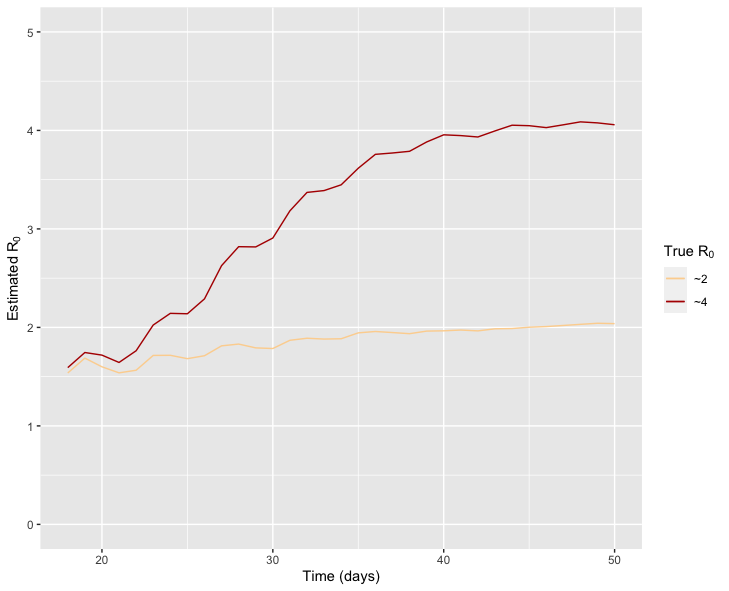}
    \caption{Effect of different underlying $R_0$ values on estimation bias for low $V(0)_{index}$ setting ($V(0)_{index}=4.5$).}
    \label{fig:r0_diff}
\end{figure}

We altered the relationship between viral load and transmission potential to assess the robustness of this estimation bias when the true $R_0$ is higher or lower (Figure \ref{fig:r0_diff}. We compared the estimation trajectory for a low $V(0)_{index}$ setting ($V(0)_{index}=4.5$) when the underlying $R_0$ is approximately two and four. We found while the early estimations are similar, the trajectories rapidly diverge and eventually stabilise around the true underlying $R_0$ value.

\section{SIR model comparison} \label{sup11}
\begin{figure}[p!]
    \centering
    \includegraphics[scale=0.4]{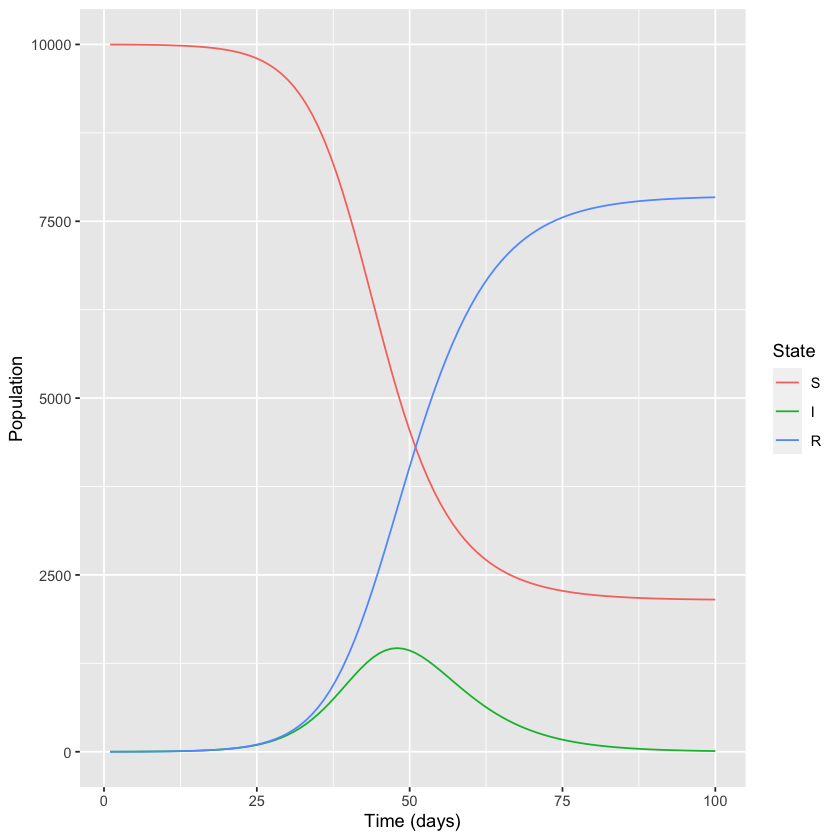}
    \caption{SIR model simulation using day 26 estimate for $R_0$ ($R_0=1.96$). We set $N=10000$, $\gamma=0.2$, and initialise each outbreak with one infected individual (i.e. $I(0)=1$). $\beta$ is set to 0.392 to align with the $R_0$ prediction.}
    \label{fig:fig10c}
\end{figure}

\begin{figure}[p!]
    \centering
    \includegraphics[scale=0.4]{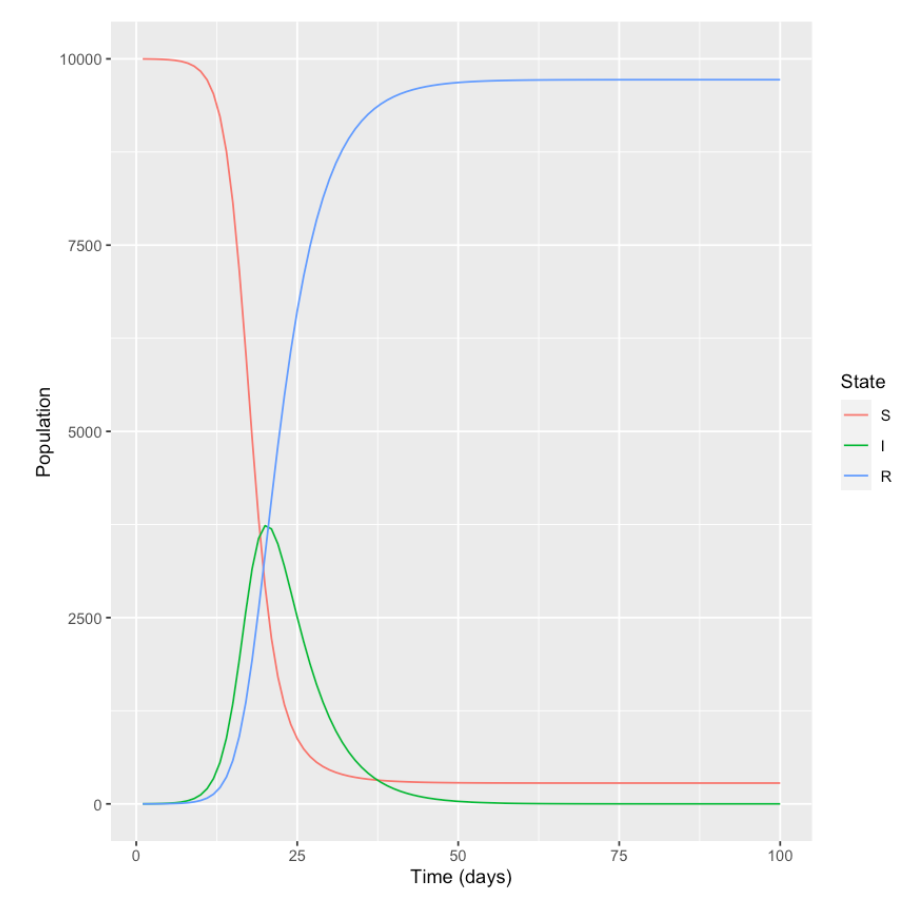}
    \caption{SIR model simulation using day 38 estimate for $R_0$ ($R_0=3.68$). We set $N=10000$, $\gamma=0.2$, and initialise each outbreak with one infected individual (i.e. $I(0)=1$). $\beta$ is set to 0.736 to align with the $R_0$ prediction.}
    \label{fig:fig10a}
\end{figure}

\begin{figure}[p!]
    \centering
    \includegraphics[scale=0.4]{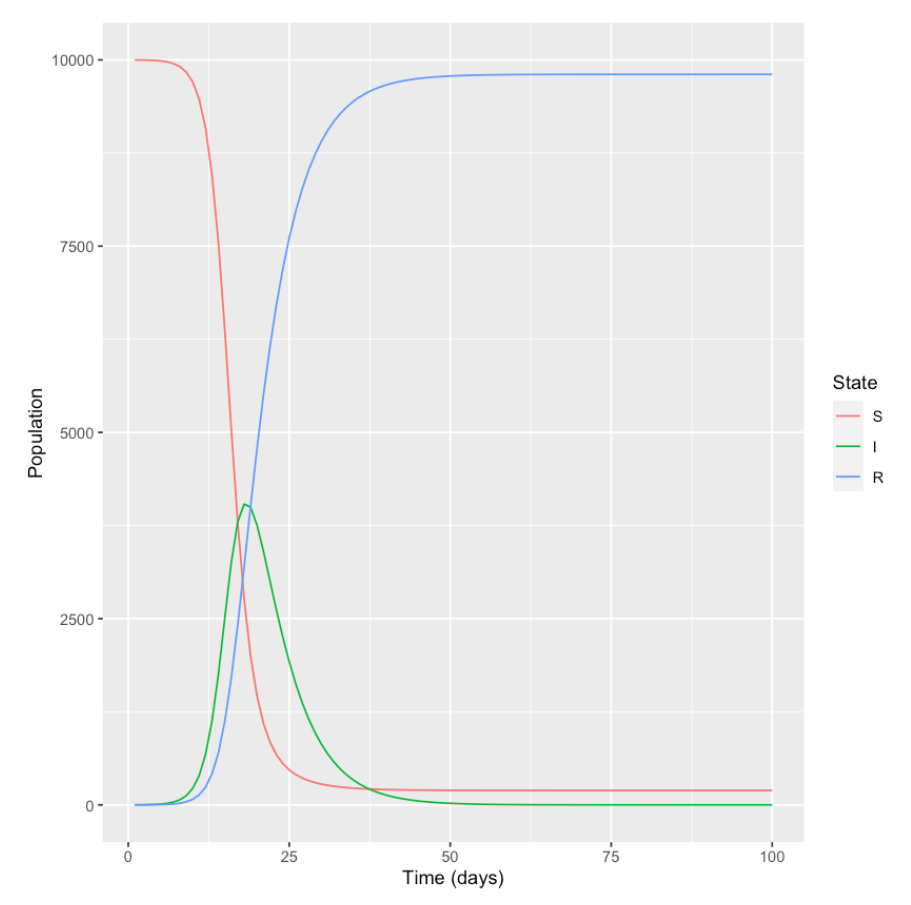}
    \caption{SIR model simulation using day 50 estimate for $R_0$ ($R_0=4.02$). We set $N=10000$, $\gamma=0.2$, and initialise each outbreak with one infected individual (i.e. $I(0)=1$). $\beta$ is set to 0.804 to align with the $R_0$ prediction.}
    \label{fig:fig10b}
\end{figure}

We used an SIR compartmental model of disease spread to simulate how the estimated $R_0$ values from Experiment 3 would affect some basic projection of disease spread. We define the system of Ordinary Differential Equations (ODEs) describing the movement of individuals between compartments as:
\begin{equation}
    \frac{dS}{dt}=-\frac{\beta S I}{N} \,,
\end{equation}
\begin{equation}  
    \frac{dI}{dt}=\frac{\beta S I}{N} - \gamma I \,,
\end{equation}
\begin{equation}
    \frac{dR}{dt}=\gamma I \,,
\end{equation}
where $S$,$I$ and $R$ refer to the susceptible, infected and recovered populations, respectively. $N$ is the total number of individuals in the population. $\beta$ is the average number of infection producing contacts per unit time. $\gamma$ is the recovery rate. 
\\\\
The basic reproduction number ($R_0$) can be defined by:
\begin{equation}
\label{R_beta_gamma}
    R_0=\frac{\beta}{\gamma} \,,
\end{equation}
We simulate different $R_0$ values by fixing the value for $\gamma$ and varying the value for $\beta$.
\\\\
We set $N=10000$, $\gamma=0.2$, and initialise each outbreak with one infected individual (i.e. $I(0)=1$). We perform three simulations using the Maximum Likelihood Estimate (MLE) of $R_0$ at 1) day 26 ($R_0=1.96$), 2) day 38 ($R_0=3.68$) and 3) day 50 ($R_0=4.02$) of the representative sample in Experiment 3. We set $\beta$ to be 1) 0.392, 2) 0.736 and 3) 0.804, as per Equation \ref{R_beta_gamma}.
\\\\
We found simulating disease spread based on the day 26 $R_0$ estimate produced a peak infected population of approximately 1465, 48 days after the outbreak was initialised (Figure \ref{fig:fig10c}). This is substantially less than the peak simulated using the day 38 and day 50 $R_0$ estimates, which produced 3735 and 4038 infected individuals, 20 and 18 days after the outbreak was initialised, respectively (Figures \ref{fig:fig10a} \& \ref{fig:fig10b}).
\\\\

\section{Sensitivity analysis: Within-host immune system parameters ($k_I$, $k_N$, $k_P$, $K_v$, $K_I$)} \label{sup12}
\begin{figure}[p!]
    \centering
    \includegraphics[scale=0.4]{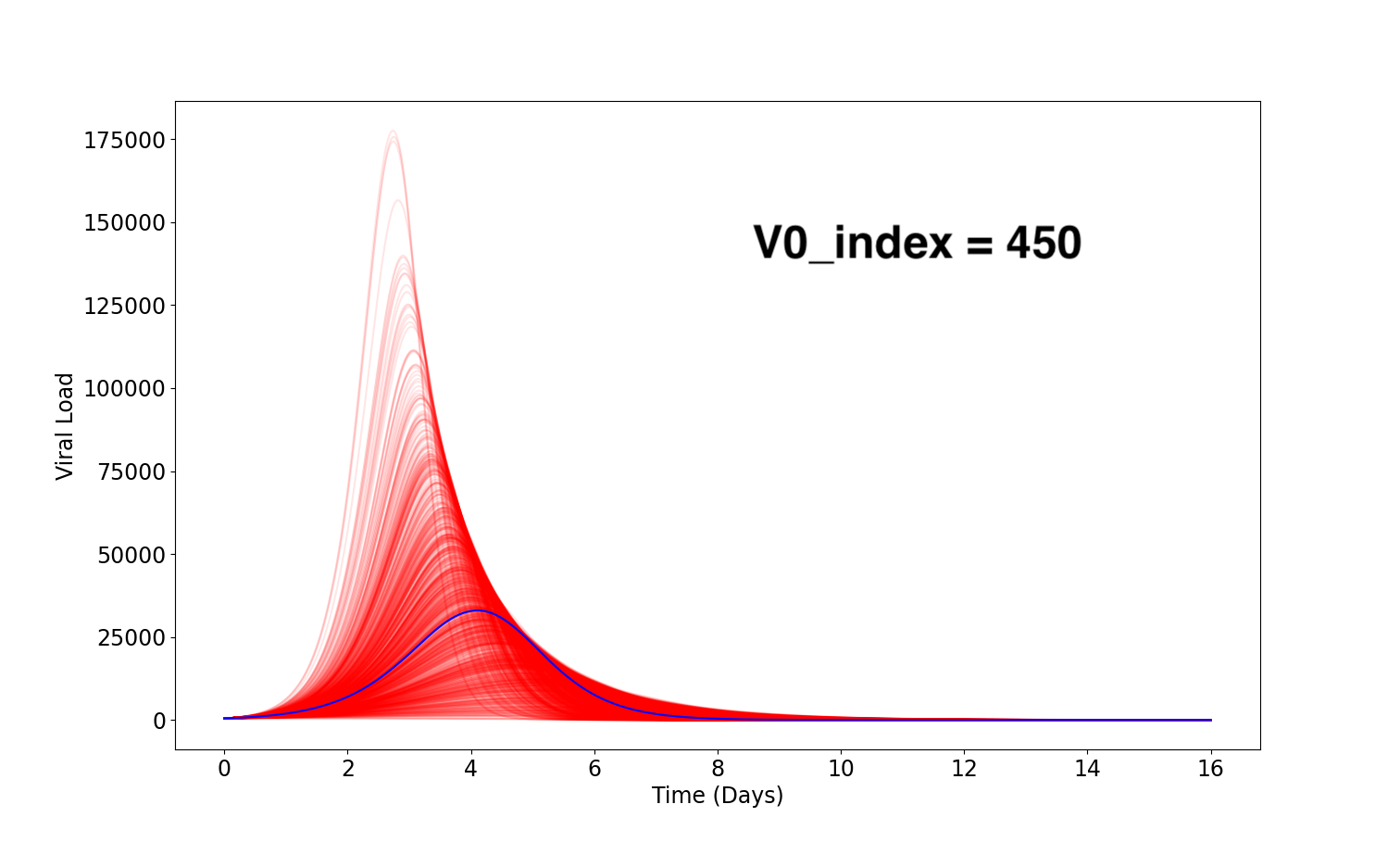}
    \caption{Sample of within-host viral load trajectories with relative variation of 0.1. Blue line represents viral load curve with no variation in within-host parameters.}
    \label{fig:sa_1}
\end{figure}

\begin{figure}[p!]
    \centering
    \includegraphics[scale=0.4]{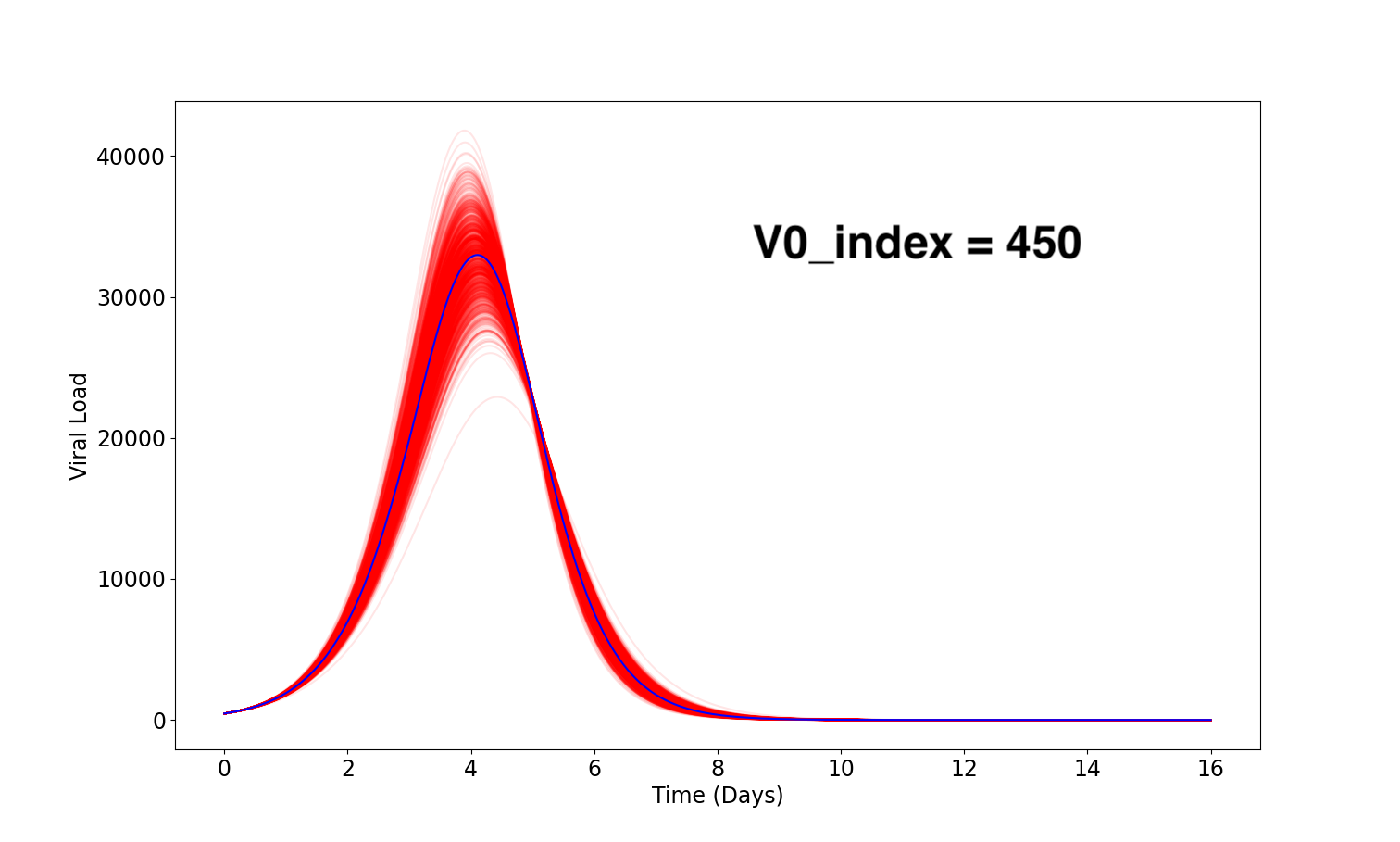}
    \caption{Sample of within-host viral load trajectories with relative variation of 0.01. Blue line represents viral load curve with no variation in within-host parameters.}
    \label{fig:sa_01}
\end{figure}

\begin{figure}[p!]
    \centering
    \includegraphics[scale=0.4]{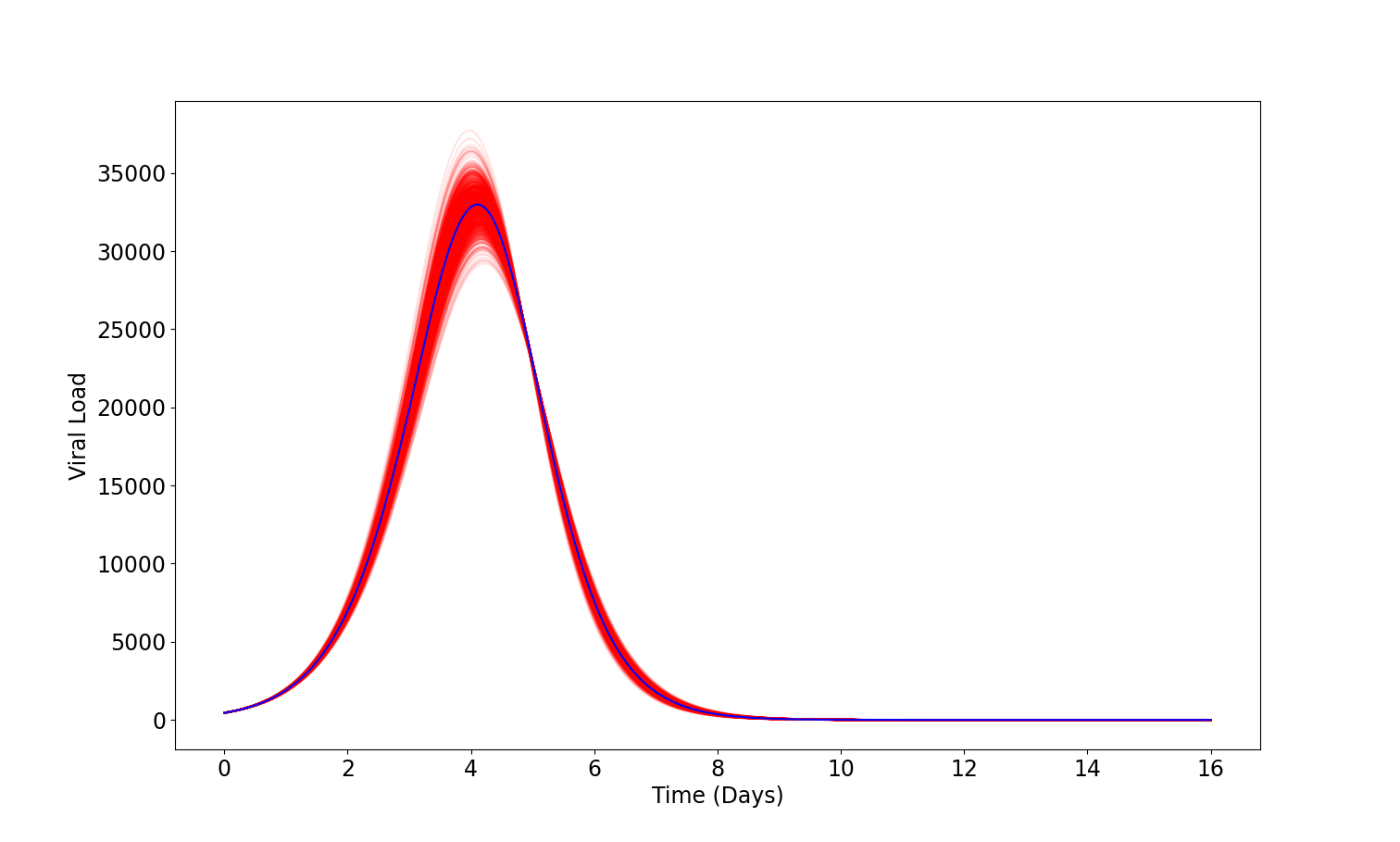}
    \caption{Sample of within-host viral load trajectories with relative variation of 0.005. Blue line represents viral load curve with no variation in within-host parameters. Note this is the relative variation used in the main analysis.}
    \label{fig:sa_005}
\end{figure}

\begin{figure}[p!]
    \centering
    \includegraphics[scale=0.4]{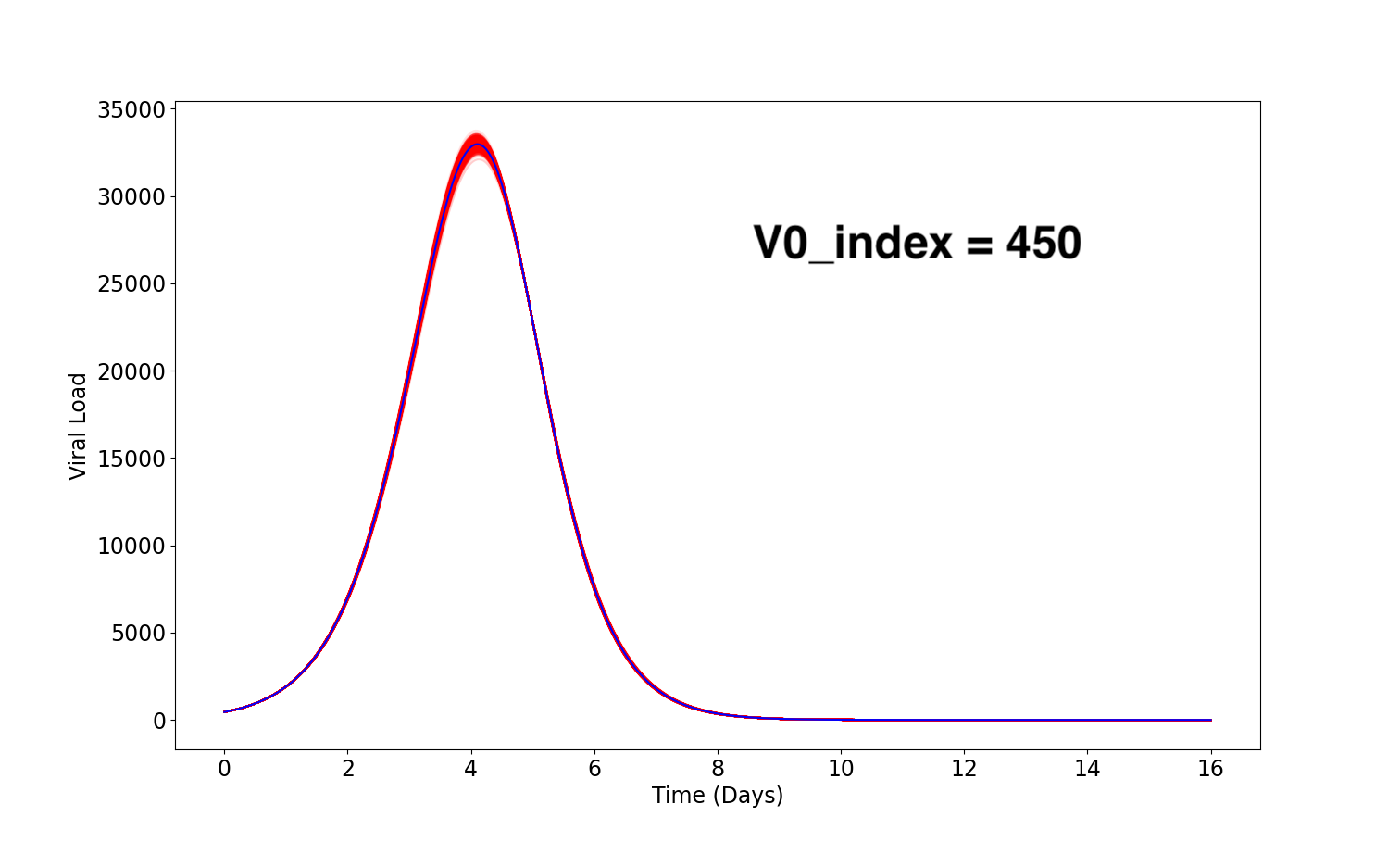}
    \caption{Sample of within-host viral load trajectories with relative variation of 0.001. Blue line represents viral load curve with no variation in within-host parameters.}
    \label{fig:sa_001}
\end{figure}

\begin{figure}[p!]
    \centering
    \includegraphics[scale=0.4]{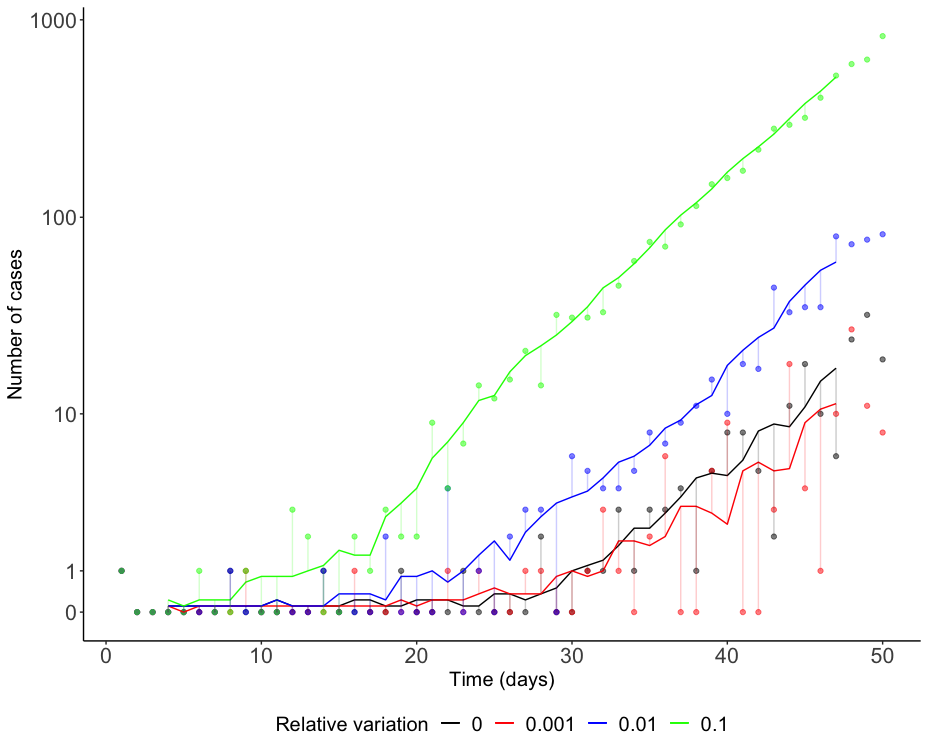}
    \caption{Initial disease incidence time series of low $V(0)_{index}$ outbreaks with alternate relative variation values. Outbreaks are seen to typically decohere faster, and subsequently grow faster when the relative variation is increased. Representative incidence trajectories shown here were chosen from a set ($N=50$) based on a defined set of outbreak summary statistics criteria. The 7-day smoothed mean (solid line) is shown alongside the raw data (points).}
    \label{fig:sa_wh}
\end{figure}

We investigated model sensitivity to the standard deviation of the Gamma distributions in the within-host parameter sampling described in the within-host model description in the main text. In the main analysis, we sample the within-host immune system parameters from Gamma distributions with a standard deviation of 0.005 multiplied by the mean of the distribution. Here, we investigated other relative variation values of 0 (i.e. no variation), 0.001, 0.01 \& 0.1. 
\\\\
First, we assessed the effect of this variation on the space of within-host viral load curves (Figures \ref{fig:sa_1}, \ref{fig:sa_01}, \ref{fig:sa_005} \& \ref{fig:sa_001}). We found the variety of viral load curve shapes observed increased as the relative variation was increased. Specifically, the size and timing of the peak viral load varied according to the relative variation used.
\\\\
Second, we assessed how this relative variation parameter affected outbreak development when outbreaks were initialised with a low $V(0)_{index}$---$V(0)_{index}=4.5$ (Figure \ref{fig:sa_wh}). We found outbreaks typically developed more rapidly when the relative variation was increased. This reflects the weakening of the correlation between case viral load dynamics induced by the increased relative variation. Specifically, as cases in the high relative variation scenarios are now exhibiting viral load curves that are less predictable according to their associated donor's viral load at transmission, we now see outbreaks that decohere more rapidly, and consequently, do not experience as long a phase of reduced growth at the outset of the outbreak. 

\end{document}